\documentclass[twocolumn,showpacs,preprintnumbers,amsmath,amssymb]{revtex4}
\usepackage{graphicx}% Include figure files
\usepackage{epsfig}  % Include figure files
\usepackage{epsf}    % Include figure files
\usepackage{dcolumn}% Align table columns on decimal point
\usepackage{bm}% bold math

\newcommand{\be}{\begin{equation}}
\newcommand{\ee}{\end{equation}}
\newcommand{\bea}{\begin{eqnarray}}
\newcommand{\eea}{\end{eqnarray}}

\def\l {\lambda}

\def\r {\rightarrow}
\def\bar {\overline}
\def\bbbar {B^0-\bar{B}{}^0}
\def\kkbar {K^0-\bar{K}{}^0}

\newcommand{\mdee}{{{\rm {\bf  m}}_d}}
\newcommand{\myew}{{{\rm {\bf m}}_u}}

\newcommand{\Dt}{\frac{d}{dt}}

\newcommand{\yd}{{Y}_d}

\def\bc  {\begin{center}}
\def\ec  {\end{center}}
\def\issue(#1,#2,#3){{\bf #1}, #2 (#3)} % AIP format
\def\opcit(#1){ {\em op. cit.}, #1}
\def\etal {\em et al.}
\def\APP(#1,#2,#3){Acta Phys.\ Polon.\ \issue(#1,#2,#3)}
\def\ARNPS(#1,#2,#3){Ann.\ Rev.\ Nucl.\ Part.\ Sci.\ \issue(#1,#2,#3)}
\def\CPC(#1,#2,#3){Comp.\ Phys.\ Comm.\ \issue(#1,#2,#3)}
\def\CIP(#1,#2,#3){Comput.\ Phys.\ \issue(#1,#2,#3)}
\def\EPJC(#1,#2,#3){Eur.\ Phys.\ J.\ C\ \issue(#1,#2,#3)}
\def\EPJD(#1,#2,#3){Eur.\ Phys.\ J. Direct\ C\ \issue(#1,#2,#3)}
\def\IEEETNS(#1,#2,#3){IEEE Trans.\ Nucl.\ Sci.\ \issue(#1,#2,#3)}
\def\IJMP(#1,#2,#3){Int.\ J.\ Mod.\ Phys. \issue(#1,#2,#3)}
\def\JHEP(#1,#2,#3){J.\ High Energy Physics \issue(#1,#2,#3)}
\def\JPG(#1,#2,#3){J.\ Phys.\ G\ \issue(#1,#2,#3)}
\def\MPL(#1,#2,#3){Mod.\ Phys.\ Lett.\ \issue(#1,#2,#3)}
\def\NP(#1,#2,#3){Nucl.\ Phys.\ \issue(#1,#2,#3)}
\def\NPB(#1,#2,#3){Nucl.\ Phys.\ B \issue(#1,#2,#3)}
\def\NIM(#1,#2,#3){Nucl.\ Instrum.\ Meth.\ \issue(#1,#2,#3)}
\def\PL(#1,#2,#3){Phys.\ Lett.\ \issue(#1,#2,#3)}
\def\PRD(#1,#2,#3){Phys.\ Rev.\ D \issue(#1,#2,#3)}
\def\PRL(#1,#2,#3){Phys.\ Rev.\ Lett.\ \issue(#1,#2,#3)}
\def\SJNP(#1,#2,#3){Sov.\ J. Nucl.\ Phys.\ \issue(#1,#2,#3)}
\def\ZPC(#1,#2,#3){Zeit.\ Phys.\ C \issue(#1,#2,#3)}

\begin{document} 
\preprint{CU-PHYSICS/13-2005}
                                                                                
\title{Rare Weak Decays and Direct Lepton Number Violating Signals  
in a Minimal R-Parity Violating Model of Neutrino Mass}

\author{Amitava Datta}
\affiliation{Department of Physics, Jadavpur University, Kolkata 700032,
India}

\author{Anirban Kundu}
\affiliation{Department of Physics, University of Calcutta,\\
92 Acharya Prafulla Chandra Road, Kolkata 700009, India}

\author{Jyoti Prasad Saha}
\thanks{Address after August 5, 2005: The Institute of Mathematical Sciences,
C.I.T. Campus, Taramani, Chennai 600113, India}
\affiliation{Department of Physics, Jadavpur University, Kolkata 700032,
India}

\author{Abhijit Samanta}
\affiliation{ Saha Institute of Nuclear Physics,
1/AF Bidhan Nagar, Kolkata 700064, India} 

\date{\today}

\begin{abstract}

Within the framework of R-parity violating minimal supergravity model, at
least  three relatively large lepton-number violating $\l'$ type
trilinear couplings at the GUT scale, not directly related to neutrino
physics, can naturally generate via renormalization group (RG) evolution
and/or CKM rotation the highly suppressed bilinear and trilinear
parameters at the weak scale
required to explain the neutrino oscillation data. The structure of the RG
equations and the CKM matrix restrict the choices of the three input
couplings to only eight possible combinations, each with its own 
distinctive experimental signature.
The relatively large input couplings may lead to spectacular low energy
signatures like rare weak decays of the $\tau$ lepton and K mesons,
direct lepton number violating decays of several sparticles, and 
unconventional decay modes (and reduced lifetime) of the lightest
neutralino, assumed to be the lightest supersymmetric particle (LSP),
all with sizable branching ratios.  Several low background signals at the
Tevatron and LHC have been suggested  and their sizes are estimated to be
at the observable level. From the particle content of the signal 
and the relative rate of different final states the input couplings at
the GUT scale, {\em i.e.}, the origin of neutrino masses and mixing angles, 
can be identified. 

\end{abstract}

\pacs{12.60.Jv, 14.60.Pq, 13.85.Rm} 
\maketitle

\section{Introduction}
%%%%%%%%%%%%%%%%%%%%%%%%%%%%%%%%%%%%%%%%%%%%%%%%%%%%%%%%%%%%
The discovery of neutrino oscillations \cite{nurev} has established beyond
doubt that there are at least two massive neutrinos. However, their masses
are  much smaller than the masses of the other known fermions.
Discovering the origin of these small masses is perhaps the most
challenging task for current high energy physics.

The see-saw mechanism which can be naturally accomodated in any grand
unified theory (GUT) is one of the most popular explanations of small
$\nu$ masses \cite{seesaw}. Unfortunately the simplest  version  of
this theory, a GUT  with a grand desert,
has very few  predictions for low energy physics apart from
$\nu$ masses and mixing angles. In particular the low energy spectrum
of such a theory is practically identical with that of the Standard Model
(SM).

Supersymmetry (SUSY) is the most elegant extension of the SM which solves
the naturalness problem that is inevitable in any non-supersymmetric GUT
\cite{susy}. There is a version of the minimal supersymmetric extension of
the SM (MSSM) which conserves lepton number $L$ and baryon number $B$ due to an
additional discrete symmetry called R-parity and defined in such a way that
all particles have R$=1$ and all sparticles have R$=-1$.  This theory is
referred to as the R-parity conserving (RPC) SUSY. However, there are
other interesting variations of the MSSM with appropriate discrete
symmetries which violate either baryon number or lepton number \cite{rpv}
but not both (so proton decay does not occur). 
These variants are known as the R-parity violating
(RPV) SUSY. In the most general version of this theory there are bilinear
and trilinear RPV terms \cite{rpv} in addition to the usual RPC terms.
 
One of the most interesting features of the $L$-violating
version of the RPV SUSY is that it can naturally explain the
experimentally measured $\nu$ masses and mixing angles \cite{rpvmnu}.
More important, as demanded by the naturalness argument,
this model is entirely governed by TeV scale physics. Thus the particle
spectrum consists of superpartners, {\em i.e.}, the sparticles,
of the particles  of the SM, having masses in the TeV scale.
The production and detection
of these sparticles at the ongoing Tevatron run II and the upcoming
Large Hadron Collider (LHC) or the International Linear Collider
(ILC) accelerator experiments can  directly  test the RPV models of
$\nu$ mass.

In the RPC SUSY the lightest supersymmetric particle (LSP) is stable.
In contrast  RPV SUSY allows the LSP  to decay, through the RPV couplings, to
lepton number violating channels. The multiplicity
of particles in any  event involving sparticle production
is, therefore, much larger on the average.
The other characteristic signatures \cite{rpv}
of RPV SUSY are single production of
sparticles, direct decays of sparticles via lepton number
violating  channels, and in particular the decay of the
lighter top squark \cite{rpv,valle,cdf,nkm,shibu}.

In addition to the above direct tests, indirect  signatures of RPV or RPC
SUSY can be obtained through rare weak processes
such as highly suppressed K, B, D, or $\tau$ decays or decays forbidden 
in the SM \cite{rpv,dkgen,jps-ak1}. The standard
signal is an abnormal enhancement of the branching ratio (BR)
 and/or significant change of
CP asymmetries compared to the SM expectations \cite{cpgen}.
Moreover the observed $\kkbar$ and $\bbbar$ mixings
put strong constraints on RPV parameters \cite{gg-arc,jps-ak2,jps-ak3}
as we shall see later. It should be emphasized that  RPC SUSY can
contribute to these processes only at the loop level, while RPV SUSY can
contribute at the tree level. The latter contributions can, therefore,
be potentially large and predict significant deviation of various
observables from the corresponding predictions in the SM.

No evidence of RPV or RPC SUSY has so far been found either through direct
or indirect methods. This leads to bounds on the parameter space of RPV
SUSY \cite{rpv,gb}. Particularly interesting are the bounds on the
trilinear RPV couplings which control the size of various direct or
indirect signatures. Some of these bounds, derived before the advent of
the highly constrained neutrino oscillation data \cite{nurev} or of precise
cosmological observations \cite{wmap5}, were remarkably weak and the
corresponding RPV couplings could be as large as allowed by the perturbative
nature of the theory!
The magnitudes of direct or indirect signatures of RPV SUSY
based on these bounds were, therefore, overestimated.

Many authors have revisited the bounds in the light of the neutrino data
\cite{abada,rpvnuosc,allanach}.  A major problem of the most general RPV
model is that the number of free parameters are even larger than the
almost unmanagably big parameter space of the RPC MSSM. In order to
make the analysis tractable additional simplifying assumptions restricting
the parameter space of the RPV sector were employed.

One approach is to consider a few
selected benchmark scenarios of $\nu$ oscillation. Each scenario
consists of a minimal set of RPV bilinear and trilinear couplings
at the weak scale, the number of parameters being just enough
to cope with the oscillation data \cite{abada}.  Upper bounds
on the parameters belonging to each set were then obtained by
a global analysis of the $\nu$ oscillation data.

In this paper we shall focus on the scenario
with three bilinear $\kappa_i$ and three $L$-violating
trilinear $\l'_{i33}$ couplings (the leptonic index $i$ will run from 1 to 3
in the entire paper). While the bilinears generate
a tree-level mass matrix for the neutrinos, the trilinears generate such a
matrix at the one loop level. The upper bounds on the $\l'$
couplings turn out to be  rather strong ($\sim 10^{-4}$)
\cite{abada}. As a result the
contributions of these couplings to most low energy processes except LSP decay
are negligible. One notable exception is the direct RPV decay of the
lighter top squark \cite{nkm,shibu}
if it happens to be the next to lightest supersymmetric particle (NLSP),
 a theoretically well-motivated scenario. This is so because the 
competing RPC decays of the top squark NLSP are also naturally suppressed
(see Section V for the details). Of course other RPV phenomena at
observable levels
can be accommodated in {\em ad hoc} models by arbitrarily adding other
relatively large couplings not directly related to $\nu$ phenomenology.
Such phenomena, however, are not correlated in any way with the 
$\nu$ sector and, hence, can throw no light on the origin of $\nu$ 
masses and mixing angles.

Another attractive approach is to assume that the fundamental theory at
some high scale, say the GUT scale $M_G$, has a small number of
parameters, of RPV or RPC type. A larger set of parameters required by
low energy phenomenology, including $\nu$ oscillation, can be generated
at the weak scale
via the renormalization group (RG) evolution of these parameters. 
One popular way is to
consider the usual boundary conditions at $M_G$ of the minimal
supergravity (mSUGRA) model along with three bilinear RPV parameters only
\cite{bilinear}. However, these models have very little predictive power
beyond LSP decay which has been studied in detail \cite{lspdk}.

Subsequently the full set of both bilinear and trilinear RPV couplings
were also included in the analysis \cite{bitri}.  A recent work
\cite{allanach} have obtained strong bounds on $\l'$ couplings from the
WMAP \cite{wmap5} constraints on the sum of $\nu$ masses.  They extended
the conventional mSUGRA boundary conditions at $M_G$ minimally by adding
only one non-zero $\l'$ couping. 

It should be noted that the conventional
GUTs do not include gravity. Thus any GUT  may be regarded as an effective
theory embedded in a more fundamental theory of gravity.
It is quite conceivable that this theory will lead to non-renormalizable
opoerators at $M_G$ suppressed by a heavy mass scale \cite{rpvgut}. When
the GUT symmetry is broken down and some heavy Higgs fields develop
vacuum expectation values (VEV),
bilinear and/or trilinear RPV interactions are generated at $M_G$. The
magnitudes of these parameters depend, among other things, on these VEVs.
Due to our rather limited  knowledge of the high scale physics at the
moment the magnitudes of the induced RPV parameters cannot be computed
from first principles. However, it is quite possible that only a small
subsets of these parameters have numerically significant magnitudes.
At the moment a meaningful question would be to ask what is the smallest 
set of RPV parameters which can account for low energy phenomenology 
including neutrino physics.
The magnitudes of these parameters can be restricted by low energy data.

In the model of ref. \cite{allanach} the single $\l'$ coupling at $M_G$
induce RPV bilinears at the weak scale via RG evolution. This in turn
induces one tree level $\nu$ mass at the weak scale incompatible with the
WMAP bound unless the input $\l'$ coupling is very tightly bounded from
above.  It turns out that almost all the 27 $\l'$ couplings have tiny
upper bounds \cite{allanach} so that the predictions for the indirect
signatures of RPV and/or the direct RPV decays of the sparticles, except
for the LSP, are not at observable levels attainable in the foreseeable
future.

However, a single $\l'$-type coupling cannot generate the full
$3 \times 3$ neutrino mass matrix at the weak scale with contributions from
both tree and one loop amplitudes, which could in principle be of the same
order of magnitude. Thus the stringent bounds of \cite{allanach}, though
highly suggestive, should not be regarded as a general conclusion following
from the RPV models of $m_{\nu}$.

In this paper we propose a novel mechanism with a minimal set of relatively 
large $\l'$ couplings, {\em not directly related} to $\nu$ masses, as inputs at
$M_G$. Thanks to flavor violating effects like non-diagonal Yukawa coupling 
matrices, this minimal set at $M_G$ can induce, via RG evolution, several new 
non-zero $\l'$ couplings in the weak or flavor basis at the weak scale 
with naturally suppressed magnitudes. 
The 
couplings in the physical or mass basis are obtained by applying appropriate 
Cabibbo-Kobayashi-Maskawa (CKM) 
rotations.  In addition several RPV bilinear terms
are generated, even though they are  zero at $M_G$
according to the chosen boundary conditions. While some of the
naturally suppressed induced RPV
parameters can take care of the $\nu$ oscillation data,
some relatively large $\l'$ couplings corresponding to the inputs at
$M_G$ exist at the weak scale with spectacular testable
consequences for low energy physics.

It should, however, be emphasized that the choices of the minimal input
sets are indeed very much restricted. This will be explained in Section III
by analyzing the structures of the RG equations and the CKM
matrix and by taking into account the experimental constraints 
currently available.
 
As an illustrative example we focus on the benchmark scenario
consisting of three  bilinear parameters $\kappa_i$, and three 
trilinear couplings $\l'_{i33}$ \cite{abada} at
the weak scale. We find that the set of input couplings for inducing this
benchmark scenario should be of the form $\l'_{ij3}$, $i= 1,2,3$ and $j=1,2$.
Any three of these six input couplings bearing different lepton indices
provide the minimal set we wish to find out.

For the purpose of illustration we have focussed on a particular benchmark
scenario of $\nu$ oscillation. One can in principle start with other
appropriately chosen minimal set of input parameters ({\em e.g.}, three
$\l$ type couplings) at $M_G$ having relatively large magnitudes and
induce the the desired benchmark scenario ({\em i.e.}, $\kappa_i$ and
$\l_{i33}$, \cite{abada}) at the weak scale if CKM like flavor violation
in the lepton sector is taken into account. Of course $\nu$
phenomenology alone cannot distinguish between various benchmark
scenarios, but the phenomenology of the input couplings, which we shall
discuss next, automatically points to the underlying model of $\nu$ 
oscillations.

The magnitudes of the input couplings can in principle be determined
by the data on $\nu$ masses and mixing angles
\cite{nurev}. Unfortunately the detailed predictions for the $\nu$ 
sector depend also on the parameters of the RPC sector (see below for 
the details). Practically no information on these parameters are available
at the moment. 

We hope that sufficient information will soon be available on 
the RPC sector from the measurement of sparticle masses and
BRs at Tevatron run II and LHC.  Moreover,  $\nu$ data will become
more precise in course of time. Only then a complete phenomenological fit of 
the model parameters to the oscillation data can be possible. 
For the time being, in order to
estimate the size of the signals predicted by our model, 
mainly for the purpose of illustration, we have
restricted the input couplings in a rough way. We require that the
magnitudes of the input couplings at $M_G$ be such that the induced 
parameters at the weak scale related to the
neutrino sector satisfy the upper bounds derived in \cite{abada} from
oscillation data. More discussions on this point can be found in Sections
II and V.  

The main thrust of this paper is to identify the minimal
sets of input couplings at $M_G$ which are still allowed to be 
relatively large in spite of the severe constraints from $\nu$ 
oscillation data. We then examine the interesting rare weak decays and 
collider signatures, including direct RPV decays of sparticles other
than the LSP or the top squark, which are allowed to be at the observable 
level in our model. 
 
In general the elements of the Yukawa coupling matrices in the up  and
the down quark sector, denoted by $Y_u$ and $Y_d$ respectively, cannot be
determined from the quark masses and the measured magnitudes of the 
elements of the CKM elements. On the other hand, these matrices appear 
explicitly in the RG equations. Quite often additional simplifying 
assumptions on the structures of the quark mass matrices are introduced
so that the elements of $Y_u$ and $Y_d$ are calculable from the 
observables. 

In the first part of our analysis, summarized in the above paragraphs, we
have assumed that the quark mass matrices have a particular texture such
that the CKM matrix is identical to the mixing matrix in the up quark
sector \cite{allanach,agashe}. If on the other hand a scenario
with mixing restricted to the down quark sector is considered, the
magnitudes of the input couplings at $M_G$ are very severely restricted by
the bounds of \cite{abada} and can hardly lead to any observable low
energy phenomenology apart from the LSP decay. This feature of the second
scenario has also been noted in \cite{allanach}, though their conclusion
was based on a model with a single neutrino mass.

We then take a more phenomenological approach and assume that some RPV
bilinears as well as $\l'$ couplings in the weak basis, not
directly related to $\nu$ physics, are generated at low energies
with relatively large magnitudes by some high scale
physics which is not necessarily of supergravity type.
Starting from a limited number of input coupling sets  at the weak 
scale, which are different in
general from the ones obtained in the first scenario, the
parameters required  to explain  the oscillation data can in principle be
generated through CKM rotations. However, such rotations will generate
other parameters as well. Using the currently available
 constraints on the elements of the CKM matrix \cite{PDG} and those on the
product RPV couplings arising from $\kkbar$ and $\bbbar$ mixing, we find that
the input couplings are so severely restricted that neither the rare
decays nor the direct lepton number violating decays of sparticles
(other than the decay of the top squark or the LSP) 
triggered by them will be observable.
However, both the top squark and the LSP decays will have distinctive
features characteristic of the second scenario which will be discussed in 
detail.

The plan of the paper is as follows. In Section II we establish the
notation, and discuss briefly the methodology of solving the relevant RG
equations and determination of the physical RPV couplings from the ones in
the weak basis generated by RG evolution in Scenario I (mixing
restricted to the up quark sector). In Section III we identify the
relatively large input trilinear RPV couplings at $M_G$ in Scenario I
which can induce at the weak scale the naturally suppressed parameters for 
neutrino oscillations
and other interesting couplings via RG evolution and CKM rotation. In
Section IV we discuss the testable predictions of our model for rare
decays. In Section V the RPV decays of the top squark, the LSP, and other
sparticles, and their collider signatures have been analyzed. In Section VI
the phenomenological model at the weak scale in Scenario II (mixing
restricted to the down quark sector) and its observable consequences are
examined. Our conclusions are summarized in the last Section.

%%%%%%%%%%%%%%%%%%%%%%%%%%%%%%%%%%%%%%%%%%%%%%%%%%%%%%%%%%%%%
\section{RG evolution and CKM rotation of the RPV couplings}

With an eye on the neutrino phenomenology, we  confine ourselves
to $L$-violating couplings only. There are three such terms in the
superpotential:
\be
W_{L\!\!\!/} =
\l_{ijk}L_i L_j E^c_k + \l'_{ijk}L_iQ_jD^c_k + \kappa_i L_i H_2
  \label{superpot}
\ee
where $L,E,Q,D$ and $H_2$ stand for doublet lepton, singlet charged
lepton, doublet quark, singlet down-type quark, and the second Higgs 
doublet
(that gives mass to the top quark) superfields, and $i,j,k=1,2,3$ are
generation indices. The first term is antisymmetric in $i$ and $j$.

Regarding the RG evolution of the RPV couplings, we follow 
\cite{allanach} and refer the reader to this paper for the details.
The anomalous dimensions of the standard Yukawa couplings as well as of
the RPV couplings, and the $\beta$-functions, including that of the
bilinear terms $\mu$ and $\kappa_i$, are given in the appendix of
\cite{allanach}, and we will not repeat them here.

For the benchmark scenarios at the weak scale we are interested in, it is 
sufficient to consider only the $\l'_{ijk}$ type couplings at $M_G$.
It is obvious from the RG equations that $\l_{ijk}$s are never generated
from $\l'_{ijk}$s through RG equations and vice versa. However, even if
the bilinear couplings $\kappa_i$ are zero at the GUT scale, they will
be generated at the weak scale mainly through a term in the RG 
equation that depends on the trilinear $\l'_{ijk}$ couplings.

The first part of the procedure is to find the RPV parameters at the weak
scale. This is done through a modified version of the code ISAJET v7.69
\cite{isajet} to which we have added the RG equations for the RPV 
parameters. The steps are outlined below in brief.

\begin{itemize}
\item
The gauge and Yukawa couplings are specified at the weak scale.
We assume that the leptonic Yukawa matrix is diagonal.
As discussed in the introduction some simplifying assumptions must be made
about the structures of the Yukawa coupling matrices $Y_u$ and $Y_d$
in the quark sector
so that their elements, which appear in the RG equations, can be
directly determined from the quark masses and mixing angles.
The quark Yukawa
matrices at the weak scale are assumed to be real and symmetric so that 
the rotation matrices in left and right sectors are the same. 
In the most general case The Cabibbo-Kobayashi-Maskawa (CKM) matrix
$V$ is given by $V=U_L^\dag D_L$ where $U_L$ and $D_L$ are the rotation
matrices for left-handed quark fields. Following earlier works
\cite{allanach,agashe} we consider two scenarios: \\
{\bf Scenario I}: $U_L=U_R=V^\dag$, $D_L=D_R=1$;\\
{\bf Scenario II}: $D_L=D_R=V$, $U_L=U_R=1$.

Subject to the above assumptions the  entries of the Yukawa matrices can 
be computed from
\begin{equation}
\begin{array}{cl}
(I) & \mdee(M_Z) = [{\mdee}]_{\mbox{diag}} (M_Z)\,, \\
& \myew(M_Z) = V_{CKM}^*\cdot [{\myew}]_{\mbox{diag}} (M_Z) \cdot 
V_{CKM}^T \,.\\
(II) & \mdee(M_Z) = V_{CKM}^*\cdot [{\mdee}]_{\mbox{diag}} (M_Z)
\cdot V_{CKM}^T\,, \\ & \myew(M_Z) = [{\myew}]_{\mbox{diag}} (M_Z) \,.\\
\end{array}
\label{models2}
\end{equation}
For simplicity, we neglect $u$ and $d$ quark masses. One can also
neglect the second generation quark masses, and all lepton masses
except that of $\tau$. The magnitude of the entries of the CKM matrix
are taken from Particle Data Group 2004 \cite{PDG}, based on the
unitarity of the matrix:
\begin{widetext}
\be
V_{CKM}=\begin{pmatrix}0.9739-0.9751 & 0.221-0.227 & 0.0029-0.0045\\
                 0.221-0.227   & 0.9730-0.9744 & 0.039-0.044\\
                 0.0048-0.014  & 0.037-0.043  & 0.999-0.9992
\end{pmatrix}.
   \label{ckm1}
\ee
\end{widetext}
For numerical calculations we consider the central value of each element,
except $V_{ub}$ and $V_{td}$.

\item
At the GUT scale the elements of $Y_u$ and $Y_d$ can be computed  
from the low energy inputs (eqs.\ (\ref{models2}) and (\ref{ckm1})) 
via the RG equations following 
an iterative procedure \cite{allanach}. The elements so computed will 
govern  the RG evolutions of the RPV parameters. 
The numerical values of $Y_u(M_G)$ and $Y_d(M_G)$ in Scenario I will  
be presented  in Section III.

Scenario II will be discussed in Section VI.

\item
The mSUGRA parameter space is specified by $m_0$, the common scalar mass,
$m_{1/2}$, the common gaugino mass, $\tan\beta$, the common trilinear
term $A_0$, and $sgn(\mu)$. We run all the gauge and Yukawa couplings upwards
until $g_1$ and $g_2$ meet; that is taken to be the unification point.
The strong coupling $g_3$ is unified by hand at that point. It is
checked whether $g_3(M_Z)$ in turn goes outside the experimental range;
it does not. The mSUGRA parameters used for the computation of various 
signals will be presented in Section V. 

\item
At the GUT scale, all mSUGRA parameters  are specified. We also specify
some $\l'$ type RPV couplings to be nonzero at $M_G$. As we have
mentioned earlier and shall further clarify in the next section, 
we need to specify at least three nonzero $\l'$s each with a different
lepton index
at the GUT scale and in the {\em weak} eigenbasis of the quark fields
in order to successfully reproduce the $\nu$ phenomenology at the weak 
scale.
\item
We run down to
the weak scale, taking all threshold effects into account. It is ensured
that the electroweak symmetry is radiatively broken determining thereby 
the magnitude of $\mu$. We also perform
some sample checks on the scalar directions of the superpotenial at
various energy scales so that no dangerous directions with unbounded
or charge-color breaking minima are encountered. Of course, this is
not a serious objection if we live in a false vacuum, but the solution
is not aesthetically  pleasing.

\item
The above running of the input RPV couplings induce other trilinear and 
bilinear couplings at the weak scale including $\kappa_i$ and $\l'_{i33}$
required for the neutrino sector.

\item
One notes the crucial role the quark mixing matrices play: without them,
no $\l'_{ijk}$ which is zero at the GUT scale can acquire a nonzero value
at the weak scale.  This can easily be checked from the RG equations as
given in \cite{allanach}.
Thus, with the CKM matrix, our minimal set of 3 $\l'$ couplings at the
GUT scale can generate all 27 couplings to be nonzero at the weak scale.
These couplings are in the {\em weak} eigenbasis and should be rotated
by the proper quark mixing matrix to get the physical couplings in the
{\em mass} eigenbasis. How this is done is shown in the next section.

\item
The RG evolution of the input $\l'$ couplings are by and large independent
of the RPC sector of the mSUGRA parameters apart from tan $\beta$ which 
relates the quark masses and the Yukawa couplings (eq.\ 2). The magnitudes 
of the 
$\kappa_i$s at the weak scale, however, also depend on the higgsino 
mass parameter $\mu$ of the RPC
sector. Usually the magnitude of this parameter is fixed by the radiative
electroweak symmetry breaking condition \cite{radesb}. This magnitude depends,
among other things, on the assumption of a common 
soft breaking scalar mass $m_0$ at
$M_G$. There are ample reasons to believe that even within the 
supergravity framework the soft breaking masses for the Higgs sector and 
the sfermion sector may be significantly different \cite{pomarol,murayama} 
at $M_G$. In such nonuniversal models the magnitude of $\mu$ may be quite
different from that in the mSUGRA model. This in turn introduces sizable
uncertainties in the magnitudes of the bilinear RPV parameters. Moreover,
both the tree level and one loop neutrino mass matrices depend on the 
mSUGRA input parameters which are poorly known at the moment.   

\item
We have studied the variation of $\l'_{ijk}$ and $\kappa_i$ with $\tan\beta$.
If $\tan\beta$ is changed from 5 to 20, the couplings which are important
in the study here, {\em viz.}, $\l'_{i13}$, $\l'_{i23}$ and $\l'_{i33}$,
change by less than 10\%. Hence our estimates of the sizes of various
collider signals (Section V) remain practically unaffected. However, $(Y_d)_
{ij}$ at $M_G$ changes significantly, and as a result there is an
appreciable change in $\kappa_i$. Thus if one attempts to fit the $\nu$
oscillation data precisely, $\tan\beta$ dependence must be taken into account.

\item
We do not attempt a detailed RPV parameter fitting using the $\nu$ 
oscillation data for reasons discussed in the last paragraph. We, however,
ensure that the input RPV parameters have the right ballpark values by
satisfying  the following criteria:

(i) The bilinear couplings $\kappa_i$s at the weak scale, related
to the elements of the tree level neutrino mass matrix,   
 should not violate the upper limits ($\sim 10^{-4}$ GeV)  \cite{abada};

(ii) The magnitude of the trilinear couplings $\l'_{i33}$,
 which are responsible for the
one loop  neutrino mass matrices, should also be bounded from above
($< 1.5\times 10^{-4}$), so that the neutrino mass squared splittings
and the mixing angles are reproduced.

 Strictly speaking the bounds of \cite{abada} were derived for a common
SUSY scale  $M_{SUSY} = 100$ GeV for all masses and mass parameters 
and should be modified when an mSUGRA mass
spectrum is considered. But since we are mainly interested in right ballpark 
values only, we ignore possible changes in the bounds, which are not
expected to be very drastic. Moreover, the bounds of \cite{abada} were
derived in a basis in which the  sneutrino VEVs vanish, whereas after
the RG evolution from the GUT scale to the weak scale we arrive at
non-zero values of the above VEVs in general. In principle a rotation
in the $H_d$-slepton space should be applied to make these VEVs zero.
However, such rotations are expected to change the couplings by factors
which are order one. We neglect such effects in our order of magnitude
estimates.

(iii) In the mass eigenbasis, all the existing constraints on the 
individual couplings and their products \cite{rpv,gb} from non-neutrino 
physics should be satisfied;

(iv) Clearly the $\nu$ oscillation data cannot be explained if all
the induced parameters 
($\kappa_i$ and  $\l'_{i33}$) have magnitudes much smaller than the 
above upper bounds. In order to estimate roughly the magnitudes of the 
input couplings, we have assumed that the magnitudes of the above 
parameters should be of the same order of magnitude of the  upper limits 
mentioned in (i) and (ii). In a forthcoming paper \cite{sujoy}, 
based on numerical calculation of the eigenvalues and 
eigenvectors of the neutrino mass matrix (including both tree level and 
one loop contributions) and comparison of the results with the measured
$\nu$ mass squared differences and mixing angles, it will be shown that
the above assumption is fairly realistic (also see Section V).     
\end{itemize}

Let us now discuss how far the $\nu$ data can be useful in constraining
the RPV scenarios under consideration. First take Scenario I.
In the mass eigenbasis, eq.\ (\ref{superpot}) reads
\be
W_{L\!\!\!/} =
{\tilde\l}'_{ijk}N_i D_jD^c_k + \bar{\l}'_{imk}E_iU_mD^c_p,
    \label{mass-spot}
\ee
with
\be
\tilde{\l}'_{ijk} = \l'_{ijk},\ \ \
\bar{\l}'_{imk} = \l'_{ijk} V^\ast_{jm}.
    \label{ckm-opt2}
\ee

In eq.\ (\ref{mass-spot})
$N_i$ and $E_i$ are respectively the neutral and charged 
components of the SU(2) doublet lepton superfield $L_i$.
Thus, in the $\nu$-sector, new couplings must be generated 
from the inputs at $M_G$ through
RG evolution alone; CKM rotations do not play any role. However, both
RG evolution and CKM rotations may combine to generate at the weak scale
interesting couplings in the  mass basis involving charged lepton fields 
(see Sections IV and V for the consequences). Also see Section VI for
Scenario II.

%The consequences for mixing restricted to the down quark sector 
%alone (scenario II) will be taken up in a later section.

\section{Choice of the input RPV parameters at $M_G$}

Our goal is to find the minimal set of trilinear RPV couplings 
at $M_G$ capable of reproducing at the weak scale
the benchmark scenario \cite{abada} for $\nu$ oscillation
under consideration. 

Substituting  for the anomalous dimension matrices in
the RGE's of the trilinear couplings, eq.\ (A23) of \cite{allanach},
it is obvious that in order to generate  a particular $\l'_{ijk}$ at the
weak scale we need a nonzero $\l'_{imn}$ (with the same lepton index)
at $M_G$.  Thus at least
three $\l'$-type couplings each bearing a different lepton index 
are required as inputs at $M_G$. The same conclusion follows if we
examine the RG equations of the $\kappa_i$s,
and require these parameters to be non-zero at the weak scale for all values
of $i$. 

Of course, one option is to consider three couplings
$\l'_{i33}$ as the inputs at the GUT scale. However, the magnitudes
of these inputs will be so small due to neutrino constraints that
no new couplings having magnitudes appropriate for observable signals
at the weak scale will be induced. Thus the only collider signals will be 
LSP decays \cite{barger} and top squark decays \cite{nkm,shibu} triggered 
by the three $\l'_{i33}$-type couplings themselves. These have,
however, been studied in detail and we shall not consider this
option further in this paper.  

Thus the desired minimal set of input parameters at $M_G$ may contain any
three of the 24 $\l'_{ijk}$-type couplings (with $j=k=3$ excluded)
each with a different lepton index.
First of all we reject $\l'_{111}$ because from neutrinoless double $\beta$
decay we get a very strong upper bound on this  ($< O(10^{-5})$). 

 The Yukawa matrices $Y_u$ and $Y_d$ at $M_G$ are needed to study the
RG equations of the RPV parameters. Following \cite{allanach} the 
elements of  the matrix $Y_u(M_G)$  are determined from the quark masses
and mixing angles. Their magnitudes in Scenario I (eq.\ (2)) are found to 
be
\be
Y_u(M_G)=\begin{pmatrix}
1.67\times 10^{-4} & 7.4\times 10^{-4} &8.4\times 10^{-5}\\
             7.36\times 10^{-4}   & 3.5\times 10^{-3} & 2.3\times 10^{-2}\\
            -9.8\times 10^{-6}& 2.3\times 10^{-3} & 5.7\times 10^{-1}
\end{pmatrix},
   \label{ckm2}
\ee
 where we have used $m_u = m_d \sim 0.0$, $m_s=0.199$ GeV,
$m_c = 1.35$ GeV, $m_b=4.83$ GeV and $m_t =175$ GeV. 
 The off-diagonal elements of $Y_d$ at $M_G$ are generated through 
mixing in the up sector as seen from  the equations given below. Starting 
from a diagonal $Y_d$ at the weak scale we have
\begin{widetext}
\bea
16\pi^2\Dt (Y_d)_{22} &=&
(Y_d)_{22}\left[-\left(\frac{7}{15}g_1^2+3 g_2^2+\frac{16}{3}g_3^2\right)
 +|(Y_u)_{2n}|^2 + 6(Y_d)_{22}^2 + 3(Y_d)_{33}^2+(Y_e)_{33}^2\right],\nonumber\\
16\pi^2\Dt (Y_d)_{33} &=&
(Y_d)_{33}\left[-\left(\frac{7}{15}g_1^2+3 g_2^2+\frac{16}{3}g_3^2\right)
+|(Y_u)_{3n}|^2+3(Y_d)_{22}^2 + 6(Y_d)_{33}^2+(Y_e)_{33}^2\right].
\eea
\end{widetext}
For the off-diagonal elements of $Y_d$ the equations are
\be
16\pi^2\Dt (Y_d)_{ij} =
(Y_d)_{jj}\times (Y_u)_{in}\times (Y_u)_{jn}.
\label{YDij}
\ee
These three equations generate the elements of $Y_d (M_G)$. The only 
numerically significant elements are
$(Y_d)_{13} = 9.5 \times 10^{-6}$ , $(Y_d)_{22} = 3.5 \times 10^{-3}$,
$(Y_d)_{23} = 2.5 \times 10^{-4}$, $(Y_d)_{32} = 1.73 \times 10^{-5}$,
and $(Y_d)_{33} = 5.64 \times 10^{-2}$; all other elements of $Y_d$ 
are negligible.

The procedure to identify the phenomenologically interesting minimal 
input sets is now outlined  below. 
\begin{itemize}
\item
As discussed in the last section, we require $\l'_{i33}$ and $\kappa_i$ 
to be of the same  order of magnitude of their upper bounds,
$1.5\times 10^{-4}$ and $1.2\times 10^{-4}$ GeV respectively at the weak 
scale. 
\item
If we consider the $\l'_{211}$ and $\l'_{311}$ couplings as inputs at
$M_G$  then from the RG equation 
\be
16\pi^2\Dt \l'_{i33} = -3 \times (\yd)_{33} (\yd)_{11} \l'_{i11}
\label{li11}
\ee
it follows that the evolution of 
$\l'_{i33}$ is controlled by the $(\yd)_{11}$ which is too small 
to  give 
$\l'_{i33}$ with the desired order of  magnitude. That is why we ignore
 $\l'_{211}$ and $\l'_{311}$ as possible inputs.

\item
If we numerically integrate the RG equations for $\l'_{i33}$ and 
$\kappa_i$ using any of $\l'_{i21}$, $\l'_{i31}$ or  $\l'_{i12}$
as inputs,
we find that the last condition can be satified only if the input 
couplings are
very large and grossly violate the existing upper bounds \cite{rpv,gb} 
from non-neutrino phenomenology. This can be understood by the following 
qualitative argument.

Suppose we integrate the RGE's of $\kappa_i$  using
the  crude approximation that all couplings other than $\kappa_i$
are constants having their respective values at $M_G$ at all
scales. We obtain
\be
{\kappa_i}^W=-1.25 (\mu)^G ({(Y_d)}_{nm})^G (\l'_{inm})^G.
   \label{kapaa}
\ee
where the superscripts $W$ and $G$ refer to the weak scale and the GUT scale 
respectively. The  constant 1.25 contains the large logarithm involving
$M_G = 2\times 10^{16}$ GeV, $M_Z = 91.1$ GeV and other 
multiplicative constants appearing in the RG equation \cite{allanach}. 
Substituting the values  of $(\mu)^G$ and $({(\yd)}_{nm})^G$  
and using $\kappa_i$ as restricted above,  we estimate  the  
values of $\l'_{inm}$ (with $n=m=3$ excluded) at $M_G$. 
We find that due to small values of the corresponding elements of  $Y_d (M_G)$, 
the parameters  $\l'_{i31}$, $\l'_{i12}$ and $\l'_{i21}$ become
too large. On the other hand $\l'_{i22}$ at $M_G$ is required to be too
small to be of any phenomenological interest, 
due to the relatively large value of $({(Y_d)}_{22})^G$. 

\item
The remaining choices  are $\l'_{il3}$ ($l=1,2$) and $\l'_{i32}$.
Using eq.\ (\ref{kapaa}) we can estimate the values of the couplings
 $\l'_{il3}$ and $\l'_{i32}$, which induces $\kappa_i^W$ in the
right ballpark.  
Integrating the RG equation for $\l'_{i33}$ using 
the same approximation as above, 
we find
\be
{\l'_{i33}}^W  \sim  2.09 {\l'_{i32}}^G \times {(Y_d)_{33}}^G
\times {(Y_d)_{32}}^G.
\label{li33a}
\ee
When the estimated values of ${\l'_{i32}}^G $ are substituted,
 ${\l'_{i33}}^W$ turns out to be $O(10^{-7})$ which is too 
small to be of any interest in  neutrino physics.
On the other hand, if the elements $(\l'_{il3})^G$ are nonzero at $M_G$,
a rough estimate of $\l'_{i33}$ at the weak scale is 
\be
{\l'_{i33}}^W  \sim  0.42 {\l'_{il3}}^G \times {(Y_u)_{3n}}^G
\times {(Y_u)_{ln}}^G.
\label{li33b}
\ee
Compairing magnitudes of the elements of the matrices  $Y_u$ and $Y_d$ at 
$M_G$, we conclude that the estimated values  
of $\l'_{il3}$ naturally reproduce values of $\l'_{i33}$ correct to the
order of magnitude estimate that is favored by $\nu$ data.  
\end{itemize}

So, at the end of the day, we have only six couplings, $\l'_{i13}$ and
$\l'_{i23}$, out of which any three with  different 
lepton indices constitute a minimal set of inputs at $M_G$.
From the numerical solutions of the
RG equations we find that $|\l'_{i13}| \leq 0.13$  and  $|\l'_{i23}|
\leq 0.26 \times 10^{-2}$ can  induce  $\kappa_i$ and
$\l'_{i33}$ with the correct order of magnitude at the weak scale.  

\section{Rare weak decays}
%%%%%%%%%%%%%%%%%%%%%%%%%%%%%%%%%%%%%

In order to study the rare weak decays in Scenario I, we start with a set
of three $\l'$-type couplings in the weak basis as inputs at the GUT scale
(Section III). For successful explanation of $\nu$    phenomenology,
these three couplings should have different lepton indices. After running
down to the weak scale, as has been outlined in Section II, we
get nonzero values for all $\l'$-type couplings, but they are still in the
weak basis. For the interaction involving $\nu$    fields no further
CKM rotation is required to get the interactions in the mass basis. For
the charged lepton interactions, on the other hand,  
we rotate the fields to the mass basis (eqs. (4) and (5)) using the
CKM elements given in eq.\ (\ref{ckm1}). The initial values of the three
nonzero couplings at $M_G$ are so chosen that at the weak scale all
$|\l'_{i33}|\leq 1.5\times 10^{-4}$. We start with the largest possible
values thus allowed, and check whether any individual induced coupling
$\bar{\l}'_{imk}$ at the weak scale and in the mass basis is in conflict
with the already existing experimental upper bound\cite{rpv,gb}. If that 
is the case,
we scale down the input couplings accordingly. Otherwise we check whether
any product of two such $\bar{\l}'$ couplings violates the corresponding 
bound,
and if that is so, we scale down the input couplings further. This
iterative process is continued till all couplings at the weak scale (and
their products too) are consistent with the experimental limits. We next
study a few specific choices of inputs at $M_G$.

\noindent{\bf {Case 1: $\l'_{i13}\not= 0$}}

For this case, to start with, $|\l'_{i13} (M_G)|$ could be at most 
$0.13$ due to oscillation constraints alone. Their magnitudes 
increase about threefold
at the weak scale. However, even the individual bounds on $\l'_{113}$ and
$\l'_{213}$ are tighter provided the squarks are not too heavy. 
The first one is $0.02(m_{\tilde b_R}/
100)$ and the second one is $0.06 
(m_{\tilde b_R}/100)$ \cite{rpv,gb} where $m_{\tilde f}$ 
is the generic sfermion mass.
The bound on  $\l'_{113}$ comes  from charge current universality (CCU) 
(as well as atomic parity violation (APV)) and that on $\l'_{213}$ 
comes from the measured ratio 
 $R_\pi\equiv \Gamma(\pi\to e\nu)/\Gamma(\pi\to \mu\nu)$ 
\cite{rpv,gb}. After taking this scaling down into account, the
RG induced couplings $|\l'_{i33}|$ relevant for the
$\nu$ sector now have acceptable order of magnitude values.  

The charged lepton couplings $|\l'_{i13}|$ in the weak basis
change to the $\bar{\l}'$ couplings in the mass basis due to CKM rotation. 
The new couplings thus generated violate the
upper bounds on the
products $|\bar \l'_{313} \ \bar \l'_{333}| 
(100/m_{\tilde l})^2< 2.0\times 10^{-3}$ 
coming from $\bbbar$ mixing and $|\bar \l'_{313} \ \bar \l'_{323}|
 (100/m_{\tilde l})^2< 2.7\times 10^{-3}$ from $\kkbar$ mixing
\cite{jps-ak3} unless the input values are further scaled down (for the 
relevent formulae see eq.(21) of \cite{jps-ak3}).   
Only {\em these} constraints, and not the ones from $\nu$
oscillation, control the ultimate upper limits of the input values of the 
$\l'$s at the GUT scale. Our final choice is 
$\l'_{113} (100/m_{\tilde b_R}) < 0.0064$, 
$\l'_{213} (100/m_{\tilde b_R}) < 0.0194$, and 
$\l'_{313} (100/m_{\tilde l})   < 0.0386$.   

Such a set of couplings generate several novel channels  for 
rare $\tau$ decays, {\em e.g.}, $\tau
\to \mu\rho$, which is forbidden in the SM. The decay 
amplitudes can be found 
in eq.\ (11) of \cite{jps-ak3}. The BR scales with $m_{\tilde b_R}^{-4}$.
In Table (\ref{tab:taubounds}) 
we show the maximum theoretical expectations vis-a-vis the
experimental upper bounds \cite{PDG} for a number of lepton flavor violating
channels, with inputs $m_{\tilde b_R}$ = 300 GeV and $m_{\tilde l}$ = 100 GeV.
It is interesting to note that the theoretical limits turn out to
be only one order of magnitude (or even less) smaller than the experimental 
data in many cases. 
%We also remind the reader that for higher 
%slepton mass the BRs scale as $(\frac{m_{\tilde l}}{100})^2$ and may be 
%still larger for heavier sleptons. 

\begin{table}
\begin{center}
\begin{tabular}{|c|c|c|c|c|}
\hline
% & \multicolumn{4}{c|}{Up Sector}
% \\ \hline
Product & Upper bound &   Process &  Predicted & Expt.  \\
coupling &at $M_W$ &&BR&limit\\
\hline
\hline
& &\ $\tau \r e \pi\;$
& 8.5$\times {10^{-8}}\;$ & 3.7$\times {10^{-6}}\;$ \\
$\;\l'_{313}\l'_{113}\;$ & 2.4$\times {10^{-3}}\;$&\ $\tau \r e \eta\;$
& 7.8$\times {10^{-8}}\;$&  8.2$\times {10^{-6}}\;$\\
 & &\ $\tau \r e \rho\;$
& 1.5$\times {10^{-7}}\;$&  2.0$\times {10^{-6}}\;$\\
\hline
\hline
 & &\ $\tau \r \mu \pi\;$
& 7.8$\times {10^{-7}}\;$ & 4.0$\times {10^{-6}}\;$\\
$\;\l'_{313}\l'_{213}\;$ & 7.2$\times {10^{-3}}\;$&\ $\tau \r \mu \eta\;$
& 7.2$\times {10^{-7}}\;$ & 9.6$\times {10^{-6}}\;$\\
 & &\ $\tau \r \mu \rho\;$
& 1.4$\times {10^{-6}}\;$ & 6.3$\times {10^{-6}}\;$\\
\hline
\end{tabular}
\caption{RPV mediated rare decays of the $\tau$ lepton with possibly large 
BRs. Since $\l'_{i13}$ is the input coupling set, there is hardly any 
difference between $\l'$ and $\bar{\l}'$ for these couplings.}
\label{tab:taubounds}
\end{center}
\end{table}

The trilinear couplings of $\nu$s with two down-type quarks do not
feel any effect of the CKM rotation, and the nonzero values are solely
generated by RG evolution. Unfortunately we do not have much interesting
phenomenology here: the decay $K^+\to\pi^+\nu\bar{\nu}$ turns out to
have a very tiny contribution, orders of magnitude smaller than the
SM prediction.

\noindent{\bf {Case 2: $\l'_{i23}\not= 0$}}

In this case the bounds are controlled by $\nu$ phenomenology only:
none of the three $\l'_{i23}$ can be more than $2.6\times 10^{-3}$ at $M_G$.
This yields unobservably tiny contributions in all rare weak
processes.

\noindent{\bf {Case 3: $\l'_{123}\not= 0, \l'_{213},\l'_{313}
\not=0$}}

 Following the identical
iterations as in Case 1, our optimum choice for GUT scale inputs come out 
to be
$\l'_{123}=0.0026$, $\l'_{213}=0.0194$, and
$\l'_{313}=0.0386$. With this, all low-energy constraints on product
couplings are satisfied (including that on $\kkbar$ mixing and $\Delta m_K$),
but this generates a large contribution to
$K^+\to\pi^+\nu\bar{\nu}$; in fact, with such a choice, the bound on
$|\bar{\l}'_{313}\bar{\l}'_{323}|$, which is $3.5\times 10^{-4}$ for 300 GeV
squarks \cite{jps-ak3}, is violated by a factor of 3. Of course, one can
scale down the input values at $M_G$, but this shows that RPV
couplings that are severely restricted by $\nu$ phenomenology can 
still generate in a correlated way a large amplitude for the 
$K^+\to\pi^+\nu\bar {\nu}$ decay, and this channel is worth watching out.
However, the BRs of various lepton flavor violating $\tau$ and $\mu$ decay
modes turn out to be much smaller than Case 1, of the order of 
$10^{-14}$ or less, and hence completely uninteresting.
%; for example,\\
%$\tau \r \mu \rho\;$ $\r$ 4.61$\times {10^{-14}}$ and 
%$\tau \r \mu \eta\;$ $\r$ 2.3$\times {10^{-14}}$  and other
%decays also have negligible Br's.

\section{Lepton number violating decays of sparticles}
\subsection{Benchmark points}
As in the last section we shall consider three choices of relatively large  
trilinear couplings as inputs at the GUT scale:\\
Case (1) \ : \ $\l'_{i13}\not= 0$ at $M_G$, \\ 
Case (2) \ : \ $\l'_{i23}\not= 0$ at $M_G$, \\ 
Case (3) \ : any combination of three non-zero couplings taken 
from Case 1 and Case 2.\\  
In all three cases the chosen trilinear couplings must carry different 
lepton indices $i$.  All the three choices can allow 
direct lepton number violating decays of the sparticles  
(sleptons, squarks and LSP) with appreciable BRs. 
More significantly the collider signals can reveal, both qualitatively and 
quantitatively, the GUT scale RPV physics related to the origin of
$\nu$    masses. 

From the upper bounds on the input couplings the precise magnitudes of the 
input couplings cannot be computed. However, it has been noted in 
\cite{sujoy} that the hierarchy $\l'_{i13} < \l'_{i23}, \l'_{i33}$ 
provides an interesting solution of the oscillation data. If we set the
magnitudes of the three input couplings in Case 1
in the ratio of their upper bounds, the desired hierarchy is maintained. 
For the purpose of illustration we
shall follow this procedure, namely, 
of setting the input RPV couplings at $M_G$ at
such values so as to reproduce, at $M_W$, the upper bounds of the respective
couplings computed with a common mass of 100 GeV for all the sfermions.
(Note that they may not be the actual upper bounds at the benchmarks
chosen by us, but one needs a standard set, valid across the different 
benchmark points, to make quantitative comparisons.) 
In case (2) the signal is qualitatively different from 
case (1). Even here, for quantitative estimates we shall set all input 
couplings equal to their upper bounds ({\em i.e.}, at 100 GeV).
Using $\l'$ scaled up by the ratio $m_{\tilde b_R}/100$ (see Section IV)
one obtains larger RPV signals. Our estimates are, therefore, conservative.

Sparticles  have  RPC decays as well. So the BR of the RPV decay 
of any sparticle  depends on the parameters of both RPV and RPC 
sectors. Computing the BRs using the above prescription on the 
input couplings, we find that sizable  BRs 
of direct RPV decays of sfermions are still allowed. However, for quantitative
estimates, we show our numbers for some favorable benchmark points. for example,
the LHC benchmark is at
\bea
&{}&m_0 = 200~{\rm GeV}, \ m_{\frac{1}{2}} = 250~{\rm GeV},\ \nonumber\\
&{}& A_0 = 0, \tan\beta = 10, sgn(\mu)= -1.
   \label{lhcbench}
\eea
From radiative electroweak symmetry breaking one gets $|{\mu}| = 345.44$ GeV.
The SUSY soft parameters are at $M_1 = 107.1$ GeV and $M_2 = 210.4$ GeV,
and the mass spectrum (in GeV) is\\
$m_{\tilde {e_L}} = m_{\tilde {\mu_L}}$ = 270.4,
$m_{\tilde {\tau_1}}$ = 220.9, $m_{\tilde {\tau_2}}$ = 271.9,
$m_{\tilde {u_L}}$ = 576.7, $m_{\tilde {d_L}}$ = 582.1,
$m_{\tilde {b_1}}$ = 526.3, $m_{\tilde {b_2}}$ = 554.7,
$m_{\tilde {\chi_1^0}}$ = 106.4, $m_{\tilde {\chi_2^0}}$ = 199.7,
$m_{\tilde {\chi_3^0}}$ = 350.4, $m_{\tilde {\chi_4^0}}$ = 361.8,
$m_{\tilde {\chi_1^+}}$ = 200.0, $m_{\tilde {\chi_2^+}}$ = 365.1,
$m_{\tilde g}$ = 634.7.

We have used three representative benchmark points 
for the Tevatron.

\noindent{\bf {Choice (1): Signals for gaugino pair production}}\\
\bea
&{}&m_0 = 200~{\rm GeV}, \ m_{\frac{1}{2}} = 145~{\rm GeV},\ \nonumber\\
&{}&A_0 = -540,\ \tan\beta = 11,\  sgn(\mu)= -1.
   \label{tevbench1}
\eea
Radiative electroweak symmetry breaking yields $|{\mu}| = -318.9$ GeV,
and SUSY soft parameters are $M_1 = 61.4$ GeV , $M_2 = 121.9$ GeV.
The SUSY mass spectrum (in GeV) is\\
$m_{\tilde {e_L}} = m_{\tilde {\mu_L}}$ = 229.7,
$m_{\tilde {\tau_1}}$ = 198.9, $m_{\tilde {\tau_2}}$ = 230.7,
$m_{\tilde {u_L}}$ = 381.3, $m_{\tilde {d_L}}$ = 389.4,
$m_{\tilde {b_1}}$ = 313.8, $m_{\tilde {b_2}}$ = 367.8,
$m_{\tilde {t_1}}$ = 153.1, $m_{\tilde {t_2}}$ = 406.4,
$m_{\tilde {\chi_1^0}}$ = 61.3, $m_{\tilde {\chi_2^0}}$ = 117.6,
$m_{\tilde {\chi_1^+}}$ = 117.6, $m_{\tilde {\chi_2^+}}$ = 337.3,
$m_{\tilde g}$ = 390.4.\\

This choice is motivated by the fact that the electroweak 
gaugino masses are slightly above the kinematic reach of LEP and should be
observable at the Tevatron.

\noindent{\bf {Choice (2): Signals from  top squark pair production }}\\
Here we use two benchmark points, namely,
\bea
{\bf (2a)}:&{}& 
m_0 = 140~{\rm GeV}, \ m_{\frac{1}{2}} = 180~{\rm GeV},\nonumber\\
&{}&A_0 = -631,\  \tan\beta = 11,\  sgn(\mu)= -1,
   \label{tevbench2a}
\eea
and
\bea
{\bf (2b)}:&{}& 
m_0 = 140~{\rm GeV}, \ m_{\frac{1}{2}} = 180~{\rm GeV},\nonumber\\
&{}&A_0 = -630,\  \tan\beta = 6,\  sgn(\mu)= 1.
   \label{tevbench2b}
\eea
For choice (2a) we have
$|{\mu}| = -381.1$ GeV,
$M_1 = 76.5$ GeV, and $M_2 = 151.32$ GeV.
The mass spectrum (in GeV) is\\
$m_{\tilde {e_L}} = m_{\tilde {\mu_L}}$ = 195.1,
$m_{\tilde {\tau_1}}$ = 136.1, $m_{\tilde {\tau_2}}$ = 200.4,
$m_{\tilde {u_L}}$ = 423.2, $m_{\tilde {d_L}}$ = 430.6,
$m_{\tilde {b_1}}$ = 346.7, $m_{\tilde {b_2}}$ = 407.7,
$m_{\tilde {t_1}}$ =  135.5, $m_{\tilde {t_2}}$ = 456.9,
$m_{\tilde {\chi_1^0}}$ = 74.3, $m_{\tilde {\chi_2^0}}$ = 140.9,
$m_{\tilde {\chi_1^+}}$ = 140.8, $m_{\tilde {\chi_2^+}}$ = 398.0, 
$m_{\tilde g}$ = 468.5.

The above choice is motivated by the requirement that the top squark be the 
NLSP. For choice (2b), $\tan\beta$ is smaller so
that the RPC loop decay is somewhat suppressed and the RPV decay with 
smaller couplings, as in Case (2), can compete with the RPC 
decay \cite{shibu}. For this case
$|{\mu}| = 385.4$ GeV,
$M_1 = 76.5$ GeV, and $M_2 = 151.3$ GeV.
The mass spectrum (in GeV) is\\
$m_{\tilde {e_L}} = m_{\tilde {\mu_L}}$ = 194.8,
$m_{\tilde {\tau_1}}$ = 152.0, $m_{\tilde {\tau_2}}$ = 198.3,
$m_{\tilde {u_L}}$ = 423.2, $m_{\tilde {d_L}}$ = 430.4,
$m_{\tilde {b_1}}$ = 354.1, $m_{\tilde {b_2}}$ = 410.7,
$m_{\tilde {t_1}}$ =  128.4, $m_{\tilde {t_2}}$ = 462.2,
$m_{\tilde {\chi_1^0}}$ = 73.4, $m_{\tilde {\chi_2^0}}$ = 138.8,
$m_{\tilde {\chi_1^+}}$ = 138.4, $m_{\tilde {\chi_2^+}}$ = 402.9,
$m_{\tilde g}$ = 468.7.

All production cross-sections and BRs in this section have been 
calculated using the software CalcHep \cite{calchep}. 

\subsection{Lightest neutralino decay}

We begin our discussions with RPV decays of the LSP, assumed to be 
$\tilde{\chi_1}^0$. These decays govern 
all RPV signals at the colliders except for the ones arising from direct 
RPV decays of sfermions.
If there are only $\l'_{i33}$-type  couplings, {\em i.e.}, only those 
required by $\nu$ physics, then the main LSP decay modes are  
$\tilde{\chi_1}^0 \rightarrow \nu_i b \bar b $ 
(assuming $m_t > m_{\tilde{\chi_1}^0}$). As has been discussed in 
\cite{shibu}
the lepton number violating nature of this decay is not obvious due to the 
missing $\nu$s. Moreover,  this
decay can be faked by RPC decays like $\tilde{\chi_2}^0\rightarrow 
\tilde{\chi_1}^0 b \bar b$. This is especially
 so if the masses  of ${\tilde{\chi_1}^0}$ and 
${\tilde{\chi_2}^0}$ do not follow the hypothesis of  
gaugino mass unification at $M_G$ and the LSP is allowed to be much 
lighter 
than the ${\tilde{\chi_2}^0}$. It should also be 
noted that the above decay may be the main decay channel of 
${\tilde{\chi_2}^0}$ if the lighter b-squark happens  to be 
significantly lighter than 
the other sfermions due to strong mixing effects 
at large tan $\beta$. 

In  our  model the LSP decay pattern is 
totally different  because of the presence of the relatively large 
$\l'_{i13}$ and/or $\l'_{i23}$ couplings. The dominant 
decay modes are $\tilde{\chi_1}^0 \rightarrow (l_i u/ \nu_i d) \bar b $
(Case 1) or $\tilde{\chi_1}^0 \rightarrow (l_i c/ \nu_i s) 
\bar b $ (Case 2). In Case 3 some lepton flavors will be accompanied 
by $u/d$ jets while the others will come in association with $c/s$ jets.
Thus the nature of the heavy flavor jets accompanying the charged lepton 
in the final state would be the qualitative feature that can 
distinguish one choice of GUT scale physics from another. 

\noindent{\bf {Case 1: $\l'_{i13}\not= 0$}}

Here our choices are $\l'_{113}=0.0064$, $\l'_{213}=0.0194$, and
$\l'_{313}=0.0386$ at $M_G$ (see Section IV), which we use for the purpose of
illustration in our numerical computations. At the weak scale the
magnitudes increase by about a factor of 3. For our model of quark mixing
(Scenario I) the physical couplings for the decays involving $\nu$s are 
the same as the ones in the flavor basis. For the decays into charged 
leptons the couplings in the two bases are practically identical as they 
are related by diagonal elements of the CKM matrix. 

These couplings generate
several novel channels for $\tilde{\chi_1}^0$ decay, tabulated in Table
\ref{tab:bounds1}. We have
also presented in this table the partial widths and the BRs of different
modes. The BRs for other values of the couplings can be computed by simple
scaling of the  partial widths.  It is to be noted that the lepton
number violating nature of the underlying interaction is obvious from the
decays involving charged leptons. Since, $\tilde{\chi_1}^0$ is a Majorana
particle it can decay into leptons of both charges with equal BRs. Thus
final states  with like sign dileptons or dileptons of different flavors
and no missing energy would be clean signals of sparticle pair production
in this model. Moreover, the
relative abundance  of final states with different leptonic compositions 
will be a strong indication of the magnitudes of underlying input 
couplings at $M_G$.

Finally from the total width computable from Table \ref{tab:bounds1}, 
we find that  
%$c\tau$ for $\tilde{\chi_1}^0$ is $\sim WHAT???.$. The LSP is, therefore, 
the LSP is going to decay inside the detector. Note that it is only a 
matter of proper scaling to
get the partial widths with a different set of input couplings. However, for
larger input couplings, the $\tilde{\chi_1}^0$ lifetiem will be correspondingly
shorter and the displaced vertex may be missed.   
  
\begin{table}
\begin{center}
\begin{tabular}{||c|c|c|c|c||}
\hline
$\tilde{\chi_1}^0$ decays & Value & Decay & Partial &   BR  \\
through               & at $M_W$ & channel & width (GeV) & at LHC \\ 
\hline
\hline
$\;\l'_{113}\;$ & 0.02 & $e^- u \bar b $ & 
$1.2\times 10^{-9}$ &$8.8\times 10^{-3}$\\
      &  &  $ \bar{\nu_e}  d \bar b $& 
$1.4\times 10^{-9}$ &$1.0\times 10^{-2}$\\
\hline
$\;\l'_{213}\;$ & 0.064 & $\mu^- u \bar b $ &
$1.3\times 10^{-8}$ &$9.0\times 10^{-2}$\\
 & & $ \bar{\nu_\mu}  d \bar b $& $1.4\times 10^{-8}$&$1.0\times 10^{-1}$\\
\hline
$\;\l'_{313}\;$ & 0.12 & $ \tau^- u \bar b $ &
$5.4\times 10^{-8}$ &$3.8\times 10^{-1}$\\
 & & $ \bar{\nu_\tau}  d \bar b $& $5.3\times 10^{-8}$ &$3.8\times 10^{-1}$\\
\hline
\hline
$\;\bar \l'_{123}\;$ & 0.0045 & $e^- c \bar b $ &
$6.3\times 10^{-11}$ &$4.5\times 10^{-4}$\\
\hline
$\;\bar \l'_{223}\;$ & 0.014 & $ \mu^- c \bar b $ &
$6.1\times 10^{-10}$ &$4.3\times 10^{-3}$\\
\hline
$\;\bar \l'_{323}\;$ & 0.03 & $ \tau^- c \bar b $ &
$3.4\times 10^{-9}$ &$2.4\times 10^{-2}$\\
\hline

\end{tabular}
\caption{BRs of LSP decays at LHC, for Case (1) parameter set. The CP
conjugate channels are implied. The last three couplings are generated by
CKM rotation and can produce only charged leptons in the final state.}
\label{tab:bounds1}
\end{center}
\end{table}

\begin{table}
\begin{center}
\begin{tabular}{||c|c|c|c|c||}
\hline
$\tilde{\chi_1}^0$ decays & Value & Decay & Partial &   BR  \\
through               & at $M_W$ & channel & width (GeV) & at LHC \\ 
\hline
\hline
$\;\l'_{123}\;$ & 0.0082 & $e^- c \bar b $ &
$2.1\times 10^{-10}$ & $1.5\times 10^{-1}$\\
& & $\bar{\nu_e}  s \bar b $& $2.3\times 10^{-10}$ &$1.6\times 10^{-1}$\\
\hline
$\;\l'_{223}\;$ & 0.0082 & $\mu^- c \bar b $ &
$2.1\times 10^{-10}$ &$1.5\times 10^{-1}$\\
& & $\bar{\nu_\mu}  s \bar b $& $2.3\times 10^{-10}$ & $1.6\times 10^{-1}$\\
\hline
$\;\l'_{323}\;$ & 0.0082 & $ \tau^- c \bar b $ &
$2.5\times 10^{-10}$ &$1.8\times 10^{-1}$\\
& & $\bar{\nu_\tau}  s \bar b $& $2.5\times 10^{-10}$ &$1.8\times 10^{-1}$\\
\hline
\hline
$\;\bar \l'_{113}\;$ & 0.0018 & $ e^- u \bar b $ &
$1.0\times 10^{-11}$ &$7.1\times 10^{-3}$\\
\hline
$\;\bar \l'_{213}\;$ & 0.0018 & $ \mu^- u \bar b $ &
$1.0\times 10^{-11}$ &$7.1\times 10^{-3}$\\
\hline
$\;\bar \l'_{313}\;$ & 0.0017 & $ \tau^- u \bar b $ &
$1.1\times 10^{-11}$ &$7.7\times 10^{-3}$\\
\hline

\end{tabular}
\caption{Same as Table \ref{tab:bounds1}, but for Case (2) parameter set.}
\label{tab:bounds2}
\end{center}
\end{table}

\begin{table}
\begin{center}
\begin{tabular}{||c|c|c|c|c||}
\hline
$\tilde{\chi_1}^0$ decays & Value & Decay & Partial &   BR  at\\
through               & at $M_W$ & channel & width (GeV) & Tevatron \\ 
\hline
\hline
 & & $e^- u \bar b $ & $2.2\times 10^{-10}$ &$1.0\times 10^{-2}$\\
$\;\l'_{113}\;$ &0.02&  & $(6.3\times 10^{-10})$ &$(5.6\times 10^{-2})$\\
 &(0.02) &$ \bar{\nu_e} d \bar b $& $2.2\times 10^{-10}$ &$1.3\times 10^{-2}$\\
 &  &                          & $(1.1\times 10^{-9})$ &$(9.8\times 10^{-2})$\\
\hline
 & & $\mu^- u \bar b $ & $1.9\times 10^{-9}$ &$9.1\times 10^{-2}$\\
$\;\l'_{213}\;$ &0.064 & & $(1.2\times 10^{-9})$ &$(1.0\times 10^{-1})$\\
 &(0.027) & $\bar{\nu_\mu}d \bar b $& $2.2\times 10^{-9}$&$1.1\times 10^{-1}$\\
 & & & $(2.0\times 10^{-9})$ &$(1.7\times 10^{-1})$\\
\hline
 & & $ \tau^- u \bar b $ & $8.0\times 10^{-9}$ &$3.9\times 10^{-1}$\\
$\;\l'_{313}\;$ & 0.12   & & $(3.2\times 10^{-9})$ &$(2.9\times 10^{-1})$\\
 & (0.03)& $ \bar{\nu_\tau}  d \bar b $& $7.9\times 10^{-9}$ &$3.8\times 10^{-1}$\\
& & & $(2.9\times 10^{-9})$ &$(2.6\times 10^{-1})$\\
\hline
\hline
$\;\bar \l'_{123}\;$ & 0.0045 & $e^- c \bar b $ &
$1.1\times 10^{-11}$ &$5.3\times 10^{-4}$\\
 & (0.0046) & & $(3.3\times 10^{-11})$ &$(3.0\times 10^{-3})$\\
\hline
$\;\bar \l'_{223}\;$ & 0.014 & $ \mu^- c \bar b $ &
$1.1\times 10^{-10}$ &$5.0\times 10^{-3}$\\
 & (0.0061) & & $(5.9\times 10^{-11})$ &$(5.2\times 10^{-3})$\\
\hline
$\;\bar \l'_{323}\;$ & 0.03 & $ \tau^- c \bar b $ &
$5.2\times 10^{-10}$ &$2.5\times 10^{-2}$\\
 & (0.0065) & & $(1.7\times 10^{-10})$ &$(1.5\times 10^{-2})$\\
\hline
\end{tabular}
\caption{Same as Table \ref{tab:bounds1}, for Case (1) and Tevatron
benchmark point 1 (point 2(a)).}
\label{tab:bounds3}
\end{center}
\end{table}

It is interesting to note that in addition to the highly suppressed
$\l'_{i33}$ couplings, required for $\nu$-physics,
this choice of input couplings also induces at $M_W$ 
$\bar{\l'}_{i23}$ couplings with reasonable magnitudes via CKM rotation,
as shown in Table \ref{tab:bounds1}. The decays triggered 
by these induced 
couplings into  final states involving charged leptons and charm have 
small BRs, but 
they  may still be observable in the clean environment of the ILC.

\noindent{\bf {Case 2: $\l'_{i23}\not= 0$}}\\
In this case none of the three input couplings $\l'_{i23}$ can be larger 
than $2.6\times 10^{-3}$ at  $M_G$ (see Section IV).
The allowed LSP decay modes, shown in Table (\ref{tab:bounds2}),
are characterized by the presence of charm and 
strange particles in the final state. These jets will be isolated 
from the $b$ jet and  hopefully can be tagged. Moreover
the strangeness quantum number of the strange particle directly emitted by 
the $\tilde{\chi_1}^0$ will be opposite 
to that emitted by the $\bar{b}$ (unless that hadronizes into a neutral
$B$ meson and oscillates). It is encouraging to note that the possibility of
$c$-jet tagging and reconstruction of $c$-flavored hadrons are being discussed
vigorously \cite{cjet}. 

This particular set of input couplings will induce  $\l'_{i13}$ with much 
reduced strengths compared to the input values in 
case 1. Thus the dominant (rare)  decay 
modes in this case would be the rare (dominant) decay modes of case 1. 
The magnitudes of the induced  couplings are given in  Table 
\ref{tab:bounds2}.

\begin{table}
\begin{center}
\begin{tabular}{||c|c|c|c|c||}
\hline
$\tilde{\chi_1}^0$ decays & Value & Decay & Partial &   BR  at\\
through               & at $M_W$ & channel & width (GeV) & Tevatron \\ 
\hline
\hline
 &  & $e^- c \bar b $ & $3.6\times 10^{-11}$ &$1.4\times 10^{-1}$\\
$\;\l'_{123}\;$ &0.0082&  & $(9.1\times 10^{-11})$ &$(8.8\times 10^{-2})$\\
 &  &  $ \bar{\nu_e} s \bar b $& $5.0\times 10^{-11}$ &$2.0\times 10^{-1}$\\
 &  &                  & $(2.2\times 10^{-10})$ &$(2.1\times 10^{-1})$\\
\hline
&  & $\mu^- c \bar b $ &
$3.6\times 10^{-11}$ &$1.4\times 10^{-1}$\\
$\;\l'_{223}\;$ &0.0082&  & $(9.1\times 10^{-11})$ &$(8.8\times 10^{-2})$\\
 &  &  $ \bar{\nu_\mu} s \bar b $& $5.0\times 10^{-11}$ &$2.0\times 10^{-1}$\\
 &  &                  & $(2.2\times 10^{-10})$ &$(2.1\times 10^{-1})$\\
\hline
&  & $ \tau^- c \bar b $ &
$3.5\times 10^{-11}$ &$1.4\times 10^{-1}$\\
$\;\l'_{323}\;$&0.0082 & & $(1.3\times 10^{-10})$ &$(1.2\times 10^{-1})$\\
 & & $ \bar{\nu_\tau}  s \bar b $& $4.2\times 10^{-11}$ &$1.7\times 10^{-1}$\\
& & & $(2.6\times 10^{-10})$ &$(2.5\times 10^{-1})$\\
\hline
\hline
$\;\bar \l'_{113}\;$ & 0.0018 & $e^- u \bar b $ &
$1.7\times 10^{-12}$ &$6.8\times 10^{-3}$\\
 &  & & $(4.4\times 10^{-12})$ &$(4.3\times 10^{-3})$\\
\hline
$\;\bar \l'_{213}\;$ & 0.0018& $ \mu^- u \bar b $ &
$1.7\times 10^{-12}$ &$6.8\times 10^{-3}$\\
 &  & & $(4.4\times 10^{-12})$ &$(4.3\times 10^{-3})$\\
\hline
$\;\bar \l'_{313}\;$ & 0.0018&$ \tau^- u \bar b $ &
$1.6\times 10^{-12}$ &$6.4\times 10^{-3}$\\
& & & $(6.2\times 10^{-12})$ &$(6.0\times 10^{-3})$\\
\hline
\end{tabular}
\caption{Same as Table \ref{tab:bounds1}, for Case (2) and Tevatron
benchmark point 1 (point (2b)).}
\label{tab:bounds4}
\end{center}
\end{table}

\noindent{\bf {Case 3}}\\

Here the three input couplings could be various combinations of the six
couplings presented in Case 1 and Case 2 where each input coupling should 
bear a different lepton index. The decay modes will now be combinations of the 
ones presented in Tables \ref{tab:bounds1} and \ref{tab:bounds2}.
Using the partial widths presented in 
these tables one can easily compute the desired BRs. 
Again the decay modes of the LSP and the relative
population of different final states may lead to  the 
underlying model.

In Tables \ref{tab:bounds3} and \ref{tab:bounds4}, we have shown the
LSP decay modes for the Tevatron benchmark points. Note that for Case 2,
we have chosen benchmark 2(b), the reason of which has been explained
earlier.

\subsection{Direct RPV decays of the sleptons}

With our choices of $\l'_{113}$ and $\l'_{213}$ in
Case 1 the direct RPV decays of the left selectron $\tilde e_L$
 and the left smuon $\tilde \mu_L$
have strongly suppressed BRs (see Table \ref{slbr}). Such decays may 
therefore occur only as rare modes. In sharp contrast the heavier
$\tau$ slepton mass eigenstate, which is dominantly a left slepton, may 
 have  sizable BRs for direct lepton 
number violating channels. However, if we scale the input 
$\l'$ couplings by $m_{\tilde b_R}/100$ for the
LHC benchmark point, even $\tilde e_L$ and $\tilde\mu_L$ will have RPV BRs
competitive with the RPC channels. In case 2 the direct RPV decays 
of all sleptons will only occur as rare modes. 

\begin{table*}
\begin{center}
\begin{tabular}{||c||c|c|c||c|c|c||}
\hline
%& && \multicolumn{1}{c|}{Up mixing}
Type of   & RPC Decay  &Partial & Branching  & RPV Decay &Partial &Branching  \\
Slepton & Channel & Width (GeV) & Ratio & Channel & Width (GeV) & Ratio \\
\hline
\hline
$\tilde e_L$ & $ \tilde{\chi_1}^0 e $ & $ 2.3 \times10^{-1}$
&$2.5\times 10^{-1}$& &&\\
& $ \tilde{\chi_2}^0 e $ & $ 2.5\times10^{-1}$& $2.7\times 10^{-1}$& 
$ \bar u b $ & $ 6.6 \times10^{-3}$& $7.2\times 10^{-3}$ \\
& $ \tilde{\chi_1}^- \nu_e $ & $ 4.3\times10^{-1}$& $4.7\times 10^{-1}$& 
$ \bar c b $ & $ 3.3 \times10^{-4}$& $3.6\times 10^{-4}$ \\
\hline
\hline
$\tilde{\mu_L}$& $ \tilde{\chi_1}^0 \mu $ &
$ 2.3 \times10^{-1}$& $2.4\times 10^{-1}$& & &\\
& $ \tilde{\chi_2}^0 \mu $ & $ 2.5 \times10^{-1}$& $2.5\times 10^{-1}$
&$ \bar u b $ & $ 6.7 \times10^{-2} $& $6.8\times 10^{-2}$ \\
& $ \tilde{\chi_1}^- \nu_\mu $ & $ 4.3 \times10^{-1}$& $4.4\times 10^{-1}$
&$ \bar c b $ & $ 3.2 \times10^{-3} $& $3.3\times 10^{-3}$ \\
\hline
\hline
$\tilde\tau_1$& $\tilde{\chi_1}^0 \tau $ & $ 6.1 \times10^{-1}$ &$9.7\times 10^{-1}$&
$ \bar u b$ &$ 1.1 \times10^{-2}$ &$1.7\times 10^{-2}$\\
& $ \tilde{\chi_2}^0 \tau $ & $ 2.6 \times10^{-3}$ & $4.1\times 10^{-3}$&
$\bar c b$ & $6.6\times 10^{-4}$ & $1.1\times 10^{-3}$ \\
& $ \tilde{\chi_1}^- \nu_\tau $ & $ 4.6 \times10^{-3}$ & $7.3\times 10^{-3}$&
&  &  \\
\hline
\hline
$\tilde{\tau_2}$ & $ \tilde{\chi_1}^0 \tau $ &
$ 2.7 \times10^{-1}$& $2.4\times 10^{-1}$&
$ \bar u b$ & $ 2.2 \times10^{-1}$& $2.0\times 10^{-1}$\\
& $ \tilde{\chi_2}^0 \tau $ & $ 2.2 \times10^{-1}$ & $2.0\times 10^{-1}$&
$\bar c b$ & $1.4\times 10^{-2}$ & $1.3\times 10^{-2}$ \\
& $ \tilde{\chi_1}^- \nu_\tau $ & $ 3.7 \times10^{-1}$ & $3.4\times 10^{-1}$&
&  &  \\
\hline
\end{tabular}
\caption{BRs of RPC and RPV decay channels of sleptons in Case (1) with the
LHC benchmark point. }
\label{slbr}
\end{center}
\end{table*}

\subsection{Direct RPV decays of the squarks}

 There is a small but non-negligible parameter space within the 
mSUGRA framework where gluinos are heavier than all  squarks.
Our choice of mSUGRA  parameters belongs to this parameter 
space. However, this
condition  may be satisfied by  squarks belonging to the 
third generation in a larger region of the parameter space.
In the mSUGRA model
the $b$ squark mass eigenstates can be lighter than the gluinos at large 
or even intermediate values of tan$\beta$ due to mixing effects. 
The more interesting case of the top squark will be discussed seperately.

Squarks lighter than the gluinos are more common if non-universal squark  
and/or gluino masses \cite{murayama,nonuni} motivated by various 
intricacies of
physics at $M_G$ are allowed. Detailed phenomenology of RPC models with
squarks lighter than the gluinos due to SUSY breaking SO(10) D-terms has
been discussed in \cite{drees}. The RPV signals can be obtained from these 
by simply adding the LSP decay.  
Since the RG equations of the RPV
$\l'$-type couplings are almost independent of the RPC parameters, 
our estimates
of the magnitudes of these couplings remain by and large unaltered even in
models beyond mSUGRA.
   
\begin{table*}
\begin{center}
\begin{tabular}{||c||c|c|c||c|c|c||}
\hline
%& && \multicolumn{1}{c|}{Up mixing}
Type of   & RPC Decay  &Partial &  Branching  & RPV Decay &Partial&Branching  \\
Squark & Channel &Width (GeV) &Ratio & Channel &Width (GeV)& Ratio \\
\hline
\hline
& $\tilde{\chi_1}^0 u $ & $6.3\times 10^{-2}$ &$1.1\times 10^{-2}$&
$ e^+ b$ & $4.6\times 10^{-3}$& $8.1\times 10^{-4}$\\
& $ \tilde{\chi_2}^0 u $ & $1.7$ & $3.1\times 10^{-1}$&&& \\
$\tilde{u_L}$ & $ \tilde{\chi_3}^0 u $ &
$9.0\times 10^{-3}$& $1.6\times 10^{-3}$&$ \mu^+ b$ &
$4.7\times 10^{-2}$& $8.3\times 10^{-3}$ \\
& $\tilde{\chi_4}^0 u $ & $8.1\times 10^{-2}$ & $1.4\times 10^{-2}$&
& &\\
& $ \tilde{\chi_1}^+ d $ & $3.5$ & $6.2\times 10^{-1}$&
$ \tau^+ b$ & $1.6\times 10^{-1}$& $2.9\times 10^{-2}$ \\
& $ \tilde{\chi_2}^+ d $ & $7.3\times 10^{-2}$& $1.3\times 10^{-2}$&
&&\\
\hline
\hline
& $ \tilde{\chi_1}^0 d $ & $8.4\times 10^{-2}$ & $1.5\times 10^{-2}$&
$ \bar{\nu_e} b$ & $4.0\times 10^{-3}$& $7.2\times 10^{-4}$\\
& $ \tilde{\chi_2}^0 d $ & $1.7$& $3.1\times 10^{-1}$&
& &\\
$\tilde{d_L}$ & $ \tilde{\chi_3}^0 d $ & $1.5\times 10^{-2}$& 
$2.7\times 10^{-3}$&$ \bar{\nu_\mu} b$ & $4.1\times 10^{-2}$ &
 $7.5\times 10^{-3}$ \\
& $\tilde{\chi_4}^0 d $ & 
$1.1\times 10^{-1}$ & $2.0\times 10^{-2}$&
& &\\
& $ \tilde{\chi_1}^- u $ & $3.1$ & $5.6\times 10^{-1}$&
$ \bar{\nu_\tau} b$ & $1.4\times 10^{-1}$ & $2.6\times 10^{-2}$ \\
& $\tilde{\chi_2}^- u $ & $2.9\times 10^{-1}$ & $5.5\times 10^{-2}$&
&&\\
\hline
\end{tabular}
\caption{BRs of RPC and RPV decay channels of up and down squarks
in Case (1) with the LHC benchmark point. }
\label{squdbr}
\end{center}
\end{table*}
%%%%%%%%%%%%%%%%%%%%%%%%%%%%%%%%%%%%%%%%%
In the LHC benchmark point and in Case (1), squarks dominantly  
decay into electroweak gauginos. We present in Table \ref{squdbr}
the BRs of RPC decays, as well as those of RPV decays through 
$\l'_{i13}$ couplings of $\tilde{u_L}$
and $\tilde{d_L}$. Only the decays into $\tau$ or $\nu_\tau$ may have
a few percent BR while the other modes are suppressed. Table \ref{sqbbr}
shows the channels for $b$ squark decay.
For the lighter mass eigenstate $\tilde b_1$ RPV modes 
have highly suppressed BRs, while the heavier mass eigenstate $\tilde b_2$ 
can decay into several RPV channels each having  a few percent BR.
It should be noted that the BRs of the RPV decays of $\tilde{b_2}$ will
increase dramatically if $m_{\tilde{b_2}} < m_t + m_{\tilde{\chi}_2}^-$. 

\subsection{Direct RPV decays of the lighter top squark}
The lighter top squark mass eigenstate $\tilde t_1$ can naturally have
a mass much
smaller than that of the other squarks. This may happen due to two
reasons. First, even if all squarks have the
same mass $m_0$ at $M_G$, the mass parameters of $\tilde{t}_L$ and
$\tilde{t}_R$ at the weak scale may be significantly smaller due to the
influence of the large top quark Yukawa coupling in the RG equations. The 
mass of the lighter eigenstate
may be further suppressed due to mixing effects in the
$\tilde{t}_L$-$\tilde{t}_R$ mass matrix, with large off-diagonal entries. i
It is quite conceivable that
$\tilde{t}_1$ happens to be the next lightest supersymmetric particle
(NLSP). If in addition the conditions
$m_{\tilde{t}_1} < m_t + m_{\tilde{\chi_1}^0}$ and
$m_{\tilde{t}_1} < m_b + m_W + m_{\tilde{\chi_1}^0}$ are satisfied, 
then the only possible RPC decay modes are\\
(i) the loop decay \cite{hikasa}:
$\tilde{t}_1 \rightarrow  c \tilde{\chi_1}^0 $\\
or (ii) the four body decay \cite{djouadi}:
$\tilde{t}_1 \rightarrow  b f \bar{f}^{\prime} \tilde{\chi_1}^0$,
where $f$ and $f'$ are light fermions. Both the decays 
occur in higher order of perturbation theory and have naturally 
suppressed widths. If the couplings of the type
$\l'_{i3j}$ are nonzero and of the order of $10^{-1}$ or $10^{-2}$,
then the RPV two body decay
\be
\tilde{t}_1 \rightarrow  l_i^+ \bar{d_j}
\ee
overwhelm the RPC decays and occur with 100\% BRs. However, If the
underlying trilinear coupling is order $10^{-3}$ or smaller then
the BR of each competing mode  may have a sizable magnitude. In fact it 
was shown in \cite{nkm} that the data from Tevatron Run I or the 
preliminary data from Run II are already sensitive to $\l'_{i3j} \sim
10^{-3}$-$10^{-4}$, although for rather small $m_{\tilde{t}_1}$. It was 
shown in \cite{shibu} that a larger range of $m_{\tilde{t}_1}$ can be 
probed at Run II with integrated luminosity $\sim 2$ fb$^{-1}$.    
%------------------

\begin{table*}
\begin{center}
\begin{tabular}{||c||c|c|c||c|c|c||}
\hline
%& && \multicolumn{1}{c|}{Up mixing}
Type of   & RPC Decay  &Partial&   Branching  & RPV Decay &Partial&Branching  \\
Squark & Channel &Width (GeV)& Ratio & Channel &Width (GeV)& Ratio \\
\hline
\hline
& $ \tilde{\chi_1}^0 b $ & $1.7\times 10^{-1}$ & $4.8\times 10^{-2}$&
$ e^- u $ & $5.6\times 10^{-4}$ & $1.6\times 10^{-4}$\\
& $\tilde{\chi_2}^0 b $ & $1.5$ & $4.1\times 10^{-1}$&
$ \mu u $ & $5.8\times 10^{-3}$ & $1.6\times 10^{-3}$ \\
$\tilde b_1$& $\tilde{\chi_3}^0 b $ &
$8.8\times 10^{-2}$& $2.5\times 10^{-2}$&
$\tau u$ & $2.0\times 10^{-2}$&$5.7\times 10^{-3}$ \\
& $ \tilde{\chi_4}^0 b $ &
$8.2\times 10^{-2}$ & $2.3\times 10^{-2}$&
$ \bar{\nu_e} d$ & $4.8\times 10^{-4}$ & $1.4\times 10^{-4}$ \\
& $ \tilde{\chi_1}^- t $ & $1.7$&$4.8\times 10^{-1}$&
$ \bar{\nu_\mu} d $ & $4.9\times 10^{-3}$ & $1.4\times 10^{-3}$ \\
& $ \tilde{\chi_2}^- t $ &does not& occur
&$ \bar{\nu_\tau} d $& $1.7\times 10^{-2}$ & $4.8\times 10^{-3}$\\
& & & & $ec$ & $2.8\times 10^{-5}$ & $7.9\times 10^{-6}$ \\
& & & & $\mu c$ & $2.8\times 10^{-4}$ & $7.9\times 10^{-5}$ \\
& & & & $\tau c$ & $1.3\times 10^{-3}$ & $3.6\times 10^{-4}$ \\
\hline
& $\tilde{\chi_1}^0 b $ & 
$2.0\times 10^{-1}$ & $1.3\times 10^{-1}$&
$ e u $ & 
$3.8\times 10^{-3}$ & $2.4\times 10^{-3}$\\
& $ \tilde{\chi_2}^0 b $ & 
$7.4\times 10^{-2}$&$4.7\times 10^{-2}$&
$ \mu u $ & 
$3.9\times 10^{-2}$&$2.5\times 10^{-2}$ \\
$\tilde b_2$& $ \tilde{\chi_3}^0 b $ &
$1.8\times 10^{-1}$& $1.1\times 10^{-1}$&$ \tau u$ &
$1.4\times 10^{-1}$&$8.8\times 10^{-2}$ \\
& $ \tilde{\chi_4}^0 b $ & $1.9\times 10^{-1}$&$1.2\times 10^{-1}$&
$ \bar{\nu_e} d$ & $3.3\times 10^{-3}$&$2.1\times 10^{-3}$ \\
& $ \tilde{\chi_1}^- t $ & $9.0\times 10^{-2}$&$5.8\times 10^{-2}$&
$ \bar{\nu_\mu} d $ & $3.4\times 10^{-2}$&$2.2\times 10^{-2}$ \\
&$ \tilde{\chi_2}^- t $ & $4.8\times 10^{-1}$&$3.1\times 10^{-1}$&
$ \bar{\nu_\tau} d $& $1.2\times 10^{-1}$&$7.6\times 10^{-2}$\\
& & & & $ec$ & $1.9\times 10^{-4}$ & $1.2\times 10^{-4}$ \\
& & & & $\mu c$ & $1.9\times 10^{-3}$ & $1.2\times 10^{-3}$ \\
& & & & $\tau c$ & $8.6\times 10^{-3}$ & $5.5\times 10^{-3}$ \\
\hline
\end{tabular}
\caption{BRs of RPC and RPV decay channels of bottom squarks
in Case (1) with the LHC benchmark point. }
\label{sqbbr}
\end{center}
\end{table*}
%\newpage

If $\l'_{i33}$, {\em i.e.}, the couplings required by the $\nu$ sector, 
are the only ones with appreciable magnituds then the main decay channels of 
$\tilde{t}_1$ would be
\be
\tilde{t}_1 \rightarrow  l_i^+ \bar{b}.
\ee
In contrast our model allows  other  possibilities, particularly
in Scenario II (Section VI),  depending on the
underlying model of CKM mixing and the input RPV couplings. First
consider  Scenario I. Here the three input couplings in the flavor
basis at $M_G$ are possible combinations of  $\l'_{i13}$ and  $\l'_{i23}$.
These input couplinngs can induce  $\l'_{i33}$ at the weak scale via the RG
evolution as we have already discussed. However, the input couplings,
depending on their magnitudes, may induce  couplings
of the type  $\bar{\l}'_{i33}$ involving charged lepton interactions via 
the CKM rotation which could be much larger
than the same  couplings relevant for the $\nu$ sector via RG evolution 
only. This is a more likely situation for input couplings with possibly 
larger magnitudes as in Case 1.

These $\bar{\l}'_{i33}$ couplings trigger top squark decay.
Maximum possible values of these couplings at the weak 
scale are:
\be
\l'_{133} = 8.0\times 10^{-4},\ \  \l'_{233} = 2.4\times 10^{-3},\ \  
\l'_{333} = 4.6\times 10^{-3}.
\ee
The RPV decay modes will still be $\tilde{t}_1 \rightarrow  l_i^+ \bar{b}$.
 The complete dominance of RPV decays, however, would strongly indicate the 
relatively large input couplings at the GUT scale. 

For the purpose of illustrating the competition between RPC and RPV decay
modes, we prefer to use somewhat scaled down values of the RPV couplings
of Case (1) [otherwise the RPV decay modes will be overwhelmingly large].
The input values as well as the BRs are presented in Table \ref{stopbr(I)}.
%%%%%%%%%%%%%%%%%%%%%%%%%%%%%%%%%%%%%%%%%%%%%%%
For Case (2), on the other hand, 
the upper limits of the  input couplings $\l'_{i23}$ are much smaller in 
magnitude ($\sim 10^{-3}$), and the CKM rotated $\bar{\l}'_{i33}$ couplings
generated by them would be naturally $\sim 10^{-4}$ or smaller. Thus the
possibility  of competition among RPC and RPV is better in this case,
see Table \ref{stopbr(II)}. We use the Tevatron benchmark values, but let
us emphasize that the stop signals will be much more prominent at the LHC.

In scenario II the situation is even more intriguing. For a discussion
we refer the reader to Section VI.

%%%%%%%%%%%%%%%%%%%%%%%%%%%%%%%%%%%%%%%%%%%%%%%%%%%
\begin{table}
\begin{center}
\begin{tabular}{||c||c|c|c||}
\hline
%& && \multicolumn{1}{c|}{Up mixing}
Type of   & Decay  &RPV coupling & Branching\\
Squark & Channel & and strength & Ratio \\
\hline
\hline
& $\tilde t_1 \rightarrow \tilde{\chi_1}^0 c $ &
------&$15.5\% $ \\
%&&&\\
& $\tilde t_1 \rightarrow  e^+ b $ &
$\bar \l'_{133}$ : $8.0\times 10^{-4}$ &$18.5\% $\\
%&&($8.0\times 10^{-4}$) &\\
$\tilde t_1$ & $\tilde t_1 \rightarrow \mu^+ b $ &
$\bar \l'_{233}$ : $1.03\times 10^{-3}$ &$30.7\% $\\
%&&($1.03\times 10^{-3}$) & \\
& $\tilde t_1 \rightarrow \tau^+ b $ &
$\bar \l'_{333}$ : $1.1\times 10^{-3}$ &$35.0\% $\\
%&&($1.1\times 10^{-3}$) &\\
\hline
\end{tabular}
\caption{The lighter top squark BR into RPC and RPV channels.
The input is as in Case (1), with couplings at the mass basis and at
$M_W$. The benchmark point is that of Tevatron (2a).}
\label{stopbr(I)}
\end{center}
\end{table}
%*********************************************
\begin{table}
\begin{center}
\begin{tabular}{||c||c|c|c||}
\hline
%& && \multicolumn{1}{c|}{Up mixing}
Type of   & Decay  &RPV coupling & Branching\\
Squark & Channel & and strength & Ratio \\
\hline
\hline
& $\tilde t_1 \rightarrow \tilde{\chi_1}^0 c $ &
------&$34.2\% $ \\
%&&&\\
& $\tilde t_1 \rightarrow  e^+ b $ &
$\bar \l'_{133}$ : $1.3\times 10^{-4}$ &$21.8\% $\\
%&&($1.3\times 10^{-4}$) &\\
$\tilde t_1$ & $\tilde t_1 \rightarrow \mu^+ b $ &
$\bar \l'_{233}$ : $1.3\times 10^{-4}$ &$21.8\% $\\
%&&($1.3\times 10^{-4}$) & \\
& $\tilde t_1 \rightarrow \tau^+ b $ &
$\bar \l'_{333}$ : $1.3\times 10^{-4}$ &$21.8\% $\\
%&&($1.3\times 10^{-4}$) &\\
\hline
\end{tabular}
\caption{The same as Table \ref{stopbr(I)}, but with Case (2) input and
Tevatron (2b) benchmark.}
\label{stopbr(II)}
\end{center}
\end{table}
%%%%%%%%%%%%%%%%%%%%%%%%%%%%%%%%%%%%%%%%%%%%%%%%%%%%%%
\begin{table*}
\begin{center}
\begin{tabular}{||c|c|c||c|c|c||}
\hline
 Channel &Decay modes of  &No.\ of&Channel&Decay modes of  &No.\ of  \\
 number  & stop pair      &events &number &stop pair       &events\\
\hline
\hline
1.& $\tilde{t_1} \r \tau^+ b $ & $3403$
&2.& $\tilde{t_1} \r \tau^+ b $ & $5647$\\
& $\tilde{t_1}^* \r  e^- \bar b $ 
&&& $\tilde{t_1}^* \r  \mu^- \bar b $ &\\
\hline
3.& $\tilde{t_1} \r \mu^+ b $ & $2985$
&4.& $\tilde{t_1} \r \tau^+ b $ &$80$ \\
%$\tilde{t_1} \r \mu^+ b $ & $1686$\\
& $\tilde{t_1}^* \r  e^- \bar b $ 
&&& $\tilde{t_1}^* \r  \tilde{\chi_1}^0  \bar c \r
e^+ b \bar c \ \bar u$ &\\
% $\tilde{t_1}^* \r  e^- \bar b $ &\\
\hline
5.& $\tilde{t_1} \r \tau^+ b $ & $413$&
6.& $\tilde{t_1} \r e^+ b $ & $75$\\
& $\tilde{t_1}^* \r  \tilde{\chi_1}^0  \bar c \r 
\tau^+ b \bar c \ \bar u $ 
&&&$\tilde{t_1}^* \r  \tilde{\chi_1}^0  \bar c \r
\mu^+ b \bar c \ \bar u $ &\\
\hline
7.&$\tilde{t_1} \r \mu^+ b $ & $125$
&8.&$\tilde{t_1} \r \mu^+ b $ & $363$\\
&$\tilde{t_1}^* \r  \tilde{\chi_1}^0  \bar c \r
\mu^+ b \bar c \ \bar u $ 
&&& $\tilde{t_1}^* \r  \tilde{\chi_1}^0  \bar c \r
\tau^+ b \bar c \ \bar u $& \\
\hline
9.& $\tilde{t_1} \r e^+ b $ & $218$
&10.& $\tilde{t_1} \r \tau^+ b $ & $142$\\
& $\tilde{t_1}^* \r  \tilde{\chi_1}^0  \bar c \r
\tau^+ b \bar c \ \bar u $ 
&&& $\tilde{t_1}^* \r  \tilde{\chi_1}^0  \bar c \r
\mu^+ b \bar c \ \bar u $ &\\
\hline
%\tilde{t_1} \r e^+ b $ & $87$
% & \\
%\tilde{t_1}^* \r  \tilde{\chi_1}^0  \bar c \r
%mu^+ b \bar c $ 
%&&&\\
%\hline

\end{tabular}
\caption{Number of events coming from stop-antistop pair production
at Tevatron for final
states involving dileptons with (i) like sign and same flavor; (ii)
like sign and different flavor; and (iii) unlike sign and different
flavor. The dilepton signal is accompanied by two jets. 
The charge conjugated states are also included. The input parameters are of
Case (1), Tevatron benchmark (2a), with $\tilde m_{t_1} = 135.5$ GeV. The
stop pair production cross-section is 2.92 pb.
We use the data of table \ref{stopbr(I)} and \ref{tab:bounds3}.
} 
\label{eventsnostop(I)}
\end{center}
\end{table*}

%%%%%%%%%%%%%%%%%%%%%%%%%%%%%%%%%%%%%%%%%%%%%%%%%%%%%%

\begin{table*}
\begin{center}
\begin{tabular}{||c|c|c||c|c|c||}
\hline
 Channel &Decay modes of  &No.\ of&Channel&Decay modes of  &No.\ of  \\
 number  & stop pair      &events &number &stop pair       &events\\
\hline
\hline
1.& $\tilde{t_1} \r \tau^+ b $ & $3482$
&2.& $\tilde{t_1} \r \tau^+ b $ & $3482$\\
& $\tilde{t_1}^* \r  e^- \bar b $ 
&&& $\tilde{t_1}^* \r  \mu^- \bar b $ &\\
\hline
3.& $\tilde{t_1} \r \mu^+ b $ & $3482$
&4.&$\tilde{t_1} \r e^+ b $ & $240$\\
% $\tilde{t_1} \r \mu^+ b $ & $1686$\\
& $\tilde{t_1}^* \r  e^- \bar b $ 
&&&$\tilde{t_1}^* \r  \tilde{\chi_1}^0  \bar c \r
\mu^+ b \bar c \ \bar c $&\\ 
%&  $\tilde{t_1}^* \r  e^- \bar b $ &\\
\hline
5.& $\tilde{t_1} \r \tau^+ b $ & $328$&
6.& $\tilde{t_1} \r \tilde{\chi_1}^0   c \r
\tau^+ b \bar c \ c $ &$31$ \\
& $\tilde{t_1}^* \r  \tilde{\chi_1}^0  \bar c \r 
\tau^+ b \bar c \ \bar c $ 
&&& $\tilde{t_1}^* \r  \tilde{\chi_1}^0  \bar c \r
\tau^+ b \bar c \ \bar c $ &\\
\hline
7.& $\tilde{t_1} \r \mu^+ b $ & $240$&
8.& $\tilde{t_1} \r \tilde{\chi_1}^0  c \r
\mu^+ b \bar c \ c$ &$17$ \\
& $\tilde{t_1}^* \r  \tilde{\chi_1}^0  \bar c \r 
\mu^+ b \bar c \ \bar c $ 
&&& $\tilde{t_1}^* \r  \tilde{\chi_1}^0  \bar c \r
\mu^+ b \bar c \ \bar c $ &\\
\hline
9.& $\tilde{t_1} \r e^+ b $ & $240$&
10.& $\tilde{t_1} \r \tilde{\chi_1}^0  c \r
e^+ b\bar c \ c$ &$17$ \\
& $\tilde{t_1}^* \r  \tilde{\chi_1}^0  \bar c \r 
e^+ b\bar c \ \bar c $ 
&&& $\tilde{t_1}^* \r  \tilde{\chi_1}^0  \bar c \r
e^+ b\bar c \ \bar c $ &\\
\hline
11.&$\tilde{t_1} \r \mu^+ b $ & $328$
&12.&$\tilde{t_1} \r e^+ b $ & $328$\\
&$\tilde{t_1}^* \r  \tilde{\chi_1}^0  \bar c \r
\tau^+ b\bar c \ \bar c $ 
&&& $\tilde{t_1}^* \r  \tilde{\chi_1}^0  \bar c \r
\tau^+ b\bar c \ \bar c $& \\
\hline
 13.&$\tilde{t_1} \r e^+ b $ & $240$
&14.& $\tilde{t_1} \r \tau^+ b $ & $240$\\
& $\tilde{t_1}^* \r  \tilde{\chi_1}^0  \bar c \r
\tau^+ b \bar c \ \bar c $ 
&&& $\tilde{t_1}^* \r  \tilde{\chi_1}^0  \bar c \r
\mu^+ b \bar c \ \bar c $ &\\
\hline
%\tilde{t_1} \r e^+ b $ & $87$
% & \\
%\tilde{t_1}^* \r  \tilde{\chi_1}^0  \bar c \r
%mu^+ b \bar c $ 
%&&\\
\hline
\end{tabular}
\caption{Same as Table \ref{eventsnostop(I)}, but for Case (2) and Tevatron
benchmark (2b), where $\tilde m_{t_1} = 128.4$ GeV, and
the stop pair production cross-section is 4.07 pb.
We use the data of table \ref{stopbr(II)} and \ref{tab:bounds4}.
} 
\label{eventsnostop(II)}
\end{center}
\end{table*}
%%%%%%%%%%%%%%%%%%%%%%%%%%%%%%%%%%%%%%%%%%%%%%

\subsection{Some interesting  signals with sizable number of events at the 
Tevatron and the LHC}

 Our aim is to propose some signals, in addition to LSP decays, which 
indicate the direct RPV decay of 
at least one sfermion. A signal arising from sfermion-antisfermion 
pair production followed by direct RPV decays of both is 
not suitable for this purpose because of obvious backgrounds.
Slepton pair production, for example, would lead to
final states with jets only (see Table.\ \ref{slbr}) which will suffer
from a huge QCD background. Squark-antisquark pairs decaying into 
isolated 
opposite sign dileptons (OSDs) of the same flavor and accompanied 
by $b \bar b$ jets may suffer from a large 
Drell-Yan background accompanied by QCD jets .
However, final states with isolated like sign dilepton (LSD) pairs or 
dileptons carrying different flavors are expected to 
provide cleaner signals. 

We discuss below some interesting signals arising from sparticle pair
production and their subsequent decay. To estimate the number of events,
we use an integrated luminosity of $10^5$ pb$^{-1}$ at LHC and $9000$ pb$^{-1}$
at Tevatron.

{\subsubsection {Signals from sfermion pair production}}
%%%%%%%%%%%%%%%%%%%%%%%%

First we consider the dilepton plus dijet signals from direct RPV decays of 
the lighter top squark pair produced at the Tevatron. Even for the 
apparently unfavorable signal involving opposite sign dileptons (OSD)
of the same flavor, the backgrounds can be suppressed by suitable 
cuts \cite{nkm,shibu}. We do not analyze this signal in this work. 
Rather, we analyze only those channels which result in either OSD of 
different flavor or LSD in the final state. We present 
in Tables \ref{eventsnostop(I)} and \ref{eventsnostop(II)} the number 
of events using the BRs in Tables \ref{tab:bounds1}, \ref{tab:bounds2},
\ref{stopbr(I)}, \ref{stopbr(II)}. The number of events in several cases 
are quite encouraging. 
The reconstruction of the masses of the two top squarks when both decays 
directly to RPV channels was discussed in \cite{shibu}. In some of
the cases discussed, we need the 
reconstruction of the invariant masses of two sets: one with a lepton and 
a $b$-jet, and one with
all other visible particles in the final state, such that the two 
reconstructed invariant masses agree with each other within certain tolerance 
limit. This establishes our signal.

%%%%%%%%%%%%%%%%%%%%%%%
Next  we focus our attention on the 
final states ${l_i}^{\pm}{l_j}^{\pm} \bar b \bar b (bb)
 + X$, where X is the unobserved junk at LHC  arising from initial states not 
involving top squarks. 
One of the isolated leptons in the like sign dilepton (LSD)
pair comes from the RPC decay of a sfermion followed by the RPV  
decay of the produced LSP, which is a Majorana fermion and decays into
leptons of both flavors with equal BRs. The other one
comes  from the direct RPV decay of the second 
member of the produced fermion-antisfermion pair.
Such final states may arise from several initial states listed in Tables
\ref{eventsnosq} and \ref{eventsnosl}.
The background to this signal is expected to be small 
due to the presence of the isolated
LSDs without any accompanying missing energy. In
principle the presence of the $bb$ or $\bar b \bar b $ pair in the final
state may also help to reduce the QCD background which always involve $b
\bar{b}$. This is possible if lepton tagging and/or kaon tagging can
distinguish between b and $\bar b$ with reasonable efficiency. One can
hope that this will be achieved since the study of CP violation in
B-decays is an important component of LHC physics.
The direct lepton number violating decay of one of the sfermions can be 
identified by reconstructing the invariant mass as in the case of 
the decaying top squarks.

From Table.\ \ref{squdbr} it is clear that the 
squarks may dominantly decay into $\tau$s via the RPV channel while decays 
to $e$ and $\mu$ may be heavily suppressed. However, the BR of the 
muonic channel is still of the order of a few percent. Combining these 
informations one can estimate that a significant number of events
of  $\tau-\mu$ type (of all charge combinations) can be seen.
We have also presented in Table \ref{eventsnosq}
(see last two rows) $\tau^\pm \mu^\pm b \bar b $ events which are 
also expected to have small backgrounds. The signal size is
already encouraging.

\begin{table*}
\begin{center}
\begin{tabular}{||c|c|c|c|c||}
\hline
Channel &Squark& $\sigma$ (pb) & Decay mode of  &No of  \\
number  & pair & at LHC        & the squarks    & events\\
\hline
\hline
1.&$\tilde{b_2} \tilde{b_2}^*$ & 0.54 &$\tilde{b_2}
 \rightarrow \tau^- u$ & 86\\
&& &$\tilde{b_2}^* \rightarrow \bar b \tilde{\chi_1}^0 
\rightarrow \tau^- u \bar b \ \bar b$& \\
\hline
2.&$\tilde{b_2} \tilde{b_2}^*$ & &$\tilde{b_2} \rightarrow \tau^- u$ & 85\\
&& &$\tilde{b_2}^* \rightarrow \bar b \tilde{\chi_2}^0
\rightarrow \bar f f \tilde{\chi_1}^0 \rightarrow \tau^- u \bar b \ \bar b
\ \bar f \ f$& \\
\hline
3.&$\tilde{b_1} \tilde{b_1}^*$ & 0.77 &
$\tilde{b_1} \rightarrow \tau^- u$ & 69\\
&& & $\tilde{b_1}^* \rightarrow \bar b \tilde{\chi_2}^0
\rightarrow \bar f \ f \tilde{\chi_1}^0 \rightarrow \tau^- u \bar b \ \bar b
\ \bar f \ f$& \\
\hline
4.&$\tilde{u_L} \tilde{u_L}^*$ & 0.61&$\tilde{u_L}^* \rightarrow \tau^- 
\bar b$ & 208  \\
& & & $\tilde{u_L} \rightarrow u \tilde{\chi_2}^0 
\rightarrow \bar f \ f \tilde{\chi_1}^0 \rightarrow \tau^- u \bar b \ \bar b
\ \bar f \ f$& \\
\hline
5.&$\tilde{b_2} \tilde{b_2}^*$ && $\tilde{b_2}^* \rightarrow \mu^+ \bar u$
&  237 \\
&& &$\tilde{b_2} \rightarrow \tau^- u $& \\
\hline
6.&$\tilde{u_L} \tilde{u_L}^*$ && $\tilde{u_L}^* \rightarrow \tau^- \bar b$ 
& 32  \\
&& &$\tilde{u_L} \rightarrow \mu^+ b $& \\
\hline
7.&$\tilde{u_L} \tilde{u_L}$ & 1.02 & $\tilde{u_L} \rightarrow \tau^+ \bar b$
&  49 \\
&& &$\tilde{u_L} \rightarrow \mu^+ b $& \\
\hline
\end{tabular}
\caption{Number of events 
arising from squark-antisquark pair production at LHC, with Case (1)
input. Charge conjugate channels are included.}
\label{eventsnosq}
\end{center}
\end{table*}

\begin{table*}
\begin{center}
\begin{tabular}{||c|c|c|c|c||}
\hline
Channel &Slepton& $\sigma$ (pb) & Decay mode of  &No of  \\
number  & pair & at LHC        & the sleptons   & events\\
\hline
\hline
1.& $\tilde{\tau_2}^* \tilde{\tau_2}$ & $5.2\times 10^{-3}$ &
$\tilde{\tau_2}^* \rightarrow u 
\bar b$ &  10 \\
&& & $\tilde{\tau_2} \rightarrow \tau^- \tilde{\chi_1}^0
\rightarrow \tau^- \ \tau^- u \bar b \ \bar b$& \\
\hline
2.& $\tilde{\tau_2}^* \tilde{\tau_2}$ && $\tilde{\tau_2}^* \rightarrow u 
\bar b$ &  8  \\
&&& $\tilde{\tau_2} \rightarrow \tau^- \tilde{\chi_2}^0
\rightarrow \bar f \ f \tilde{\chi_1}^0 \rightarrow \tau^- \ 
\tau^- u \bar b \ \bar b \ \bar f f$& \\
\hline
\end{tabular}
\caption{Same as Table \ref{eventsnosq}, for stau pair production.}
\label{eventsnosl}
\end{center}
\end{table*}

The production cross-section of slepton pairs is small at LHC. In spite 
of this, the signal may be important if the signal from the squarks are 
depleted due to the presence of gluinos lighter than the squarks. 
The LSD signal from $\tau$ slepton pair production has apparently a
small size, but this may be enhanced using larger values of $\l'$ which
are still allowed (so this is in some sense a conservative estimate).

\subsubsection{Signals from gaugino pair production}
The $\tau^+\tau^+bb$ signals from gaugino pair production
at LHC and Tevatron are presented in Table \ref{enogalhc(i)} for Case(1)
and in Table \ref{enogalhc(ii)} for Case (2) input values.
Sizable number of events are expected from $\tilde{\chi_2}^0 \tilde{\chi_1}^+$
and $\tilde{\chi_1}^+ \tilde{\chi_1}^-$ pairs involving LSDs of a 
particular flavor.

We conclude this section by reiterating that all the input $\l'$s at 
$M_G$ can in principle be determined by a fit to the oscillation data
if at least a partial information about the RPC sector is available 
from collider experiments. This in turn will firmly predict the particle
content of different final states as well as the relative abundance 
of various  final states. The observation of these exciting signals will
reveal the GUT scale physics lying at the origin of $\nu$ masses and
mixing angles.     

%%%%%%%%%%%%%%%%%%%%%%%%%%%%%%%%%%%%%%%%%%%%%%%%%%%%%%%%%%%%
\begin{table*}
\begin{center}
\begin{tabular}{||c|c|c|c|c||}
\hline
Channel &Gaugino & $\sigma$ (pb) at LHC &Decay channel of  &No of events  \\
number  &pair    & (Tevatron)           & the gauginos     &at LHC (Tevatron)\\
\hline
\hline
1.&$\tilde{\chi_1}^0 \tilde{\chi_1}^0$ & $9.3\times 10^{-3}$ &
$\tilde{\chi_1}^0 \rightarrow 
\tau^- u  \ \bar b$ & 62\\
&&& $\tilde{\chi_1}^0\rightarrow \tau^- u  \ 
\bar b$&\\ 
\hline
2.&$\tilde{\chi_2}^0 \tilde{\chi_2}^0$ & $2.1\times 10^{-2}$ &
$\tilde{\chi_2}^0  \rightarrow \bar f \ f \tilde{\chi_1}^0\rightarrow\tau^- u 
\bar b \ \bar f  \ f $ & 130\\
&& & $\tilde{\chi_2}^0 
\rightarrow \bar f \ f \tilde{\chi_1}^0\rightarrow\tau^- u \bar b \ \bar f \ 
f$& \\
\hline
3.&$\tilde{\chi_2}^0 \tilde{\chi_1}^+$ &$5.3\times 10^{-1}$&
$\tilde{\chi_2}^0
\rightarrow \bar f \ f \tilde{\chi_1}^0\rightarrow\tau^- u \bar b \ 
\bar f \ f$&3350\\
&&($2.8\times 10^{-1}$) &$\tilde{\chi_1}^+
\rightarrow \bar f_1 \ f_2 \tilde{\chi_1}^0\rightarrow\tau^- u \ \bar b \
\bar f_1 \ f_2$ & (186) \\
\hline
%\hline
4.&$\tilde{\chi_1}^+ \tilde{\chi_1}^-$ &$4.7\times 10^{-1}$&
$\tilde{\chi_1}^+
\rightarrow \bar f_1 \ f_2 \tilde{\chi_1}^0\rightarrow\tau^- u \bar b 
\ \bar f_1 \ f_2$&
3100\\
&&( $3.2\times 10^{-1}$) & $\tilde{\chi_1}^-
\rightarrow f_1 \ \bar f_2 \tilde{\chi_1}^0\rightarrow\tau^- u \bar b
\ f_1 \ \bar f_2$&(214)\\
\hline
5.&$\tilde{\chi_2}^+ \tilde{\chi_2}^-$ &$2.3\times 10^{-2}$&
$\tilde{\chi_2}^+
\rightarrow \bar f_1 \ f_2 \tilde{\chi_1}^0\rightarrow\tau^- u \bar b
\ \bar f_1 \ f_2$&
152\\
&& & $\tilde{\chi_2}^-
\rightarrow f_1 \bar f_2 \tilde{\chi_1}^0\rightarrow\tau^- u \bar b
\ f_1 \ \bar f_2$& \\
\hline
\end{tabular}
\caption{Number of events from gaugino pair production for Case(1) input
at LHC (Tevatron benchmark (1)) with the final
state involving $\tau^- \tau^- \bar b \bar b + X$ (and charge 
conjugate channels).}
\label{enogalhc(i)}
\end{center}
\end{table*}
%**********************************************************

\begin{table*}
\begin{center}
\begin{tabular}{||c|c|c|c|c||}
\hline
Channel &Gaugino & $\sigma$ (pb) at LHC &Decay channel of  &No of events  \\
number  &pair    & (Tevatron)           & the gauginos     &at LHC (Tevatron)\\
\hline
\hline
1.&$\tilde{\chi_2}^0 \tilde{\chi_2}^0$ & $2.1\times 10^{-2}$&
$\tilde{\chi_2}^0  
\rightarrow \bar f \ f \tilde{\chi_1}^0\rightarrow l_i^- c \bar b  \ \bar f \ f
$ & 54\\
&& & $\tilde{\chi_2}^0 
\rightarrow \bar f \ f \tilde{\chi_1}^0\rightarrow l_i^- c \bar b \ \bar f 
\ f$& \\
\hline
%\hline
2.&$\tilde{\chi_2}^0 \tilde{\chi_1}^+$ &$5.3\times 10^{-1}$&
$\tilde{\chi_2}^0
\rightarrow \bar f \ f \tilde{\chi_1}^0\rightarrow l_i^- c \bar b \ 
\bar f \ f$&
1500\\
&&(2.9 $\times 10^{-1}$) & $\tilde{\chi_1}^+
\rightarrow \bar f_1 \ f_2 \tilde{\chi_1}^0\rightarrow l_i^- c \bar b
\ \bar f_1 \ f_2$&(73) \\
\hline
3.&$\tilde{\chi_1}^+ \tilde{\chi_1}^-$ &$4.7\times 10^{-1}$&
$\tilde{\chi_1}^+
\rightarrow \bar f_1 \ f_2 \tilde{\chi_1}^0\rightarrow l_i^- c \bar b 
 \ \bar f_1 \ f_2$&
1318\\
&&(3.2 $\times 10^{-1}$) & $\tilde{\chi_1}^-
\rightarrow f_1 \ \bar f_2 \tilde{\chi_1}^0\rightarrow l_i^- c \bar b
\ f_1 \ \bar f_2$&(85) \\
\hline
4.&$\tilde{\chi_2}^+ \tilde{\chi_2}^-$ &$2.3\times 10^{-2}$&
$\tilde{\chi_2}^+
\rightarrow \bar f_1  \ f_2 \tilde{\chi_1}^0\rightarrow l_i^- c \bar b
\ \bar f_1 \ f_2$&
64\\
&& & $\tilde{\chi_2}^-
\rightarrow f_1 \ \bar f_2 \tilde{\chi_1}^0\rightarrow l_i^- c \bar b
\ f_1 \ \bar f_2$& \\
\hline
\end{tabular}
\caption{Same as Table \ref{enogalhc(i)}, but for Case (2) input,
and benchmark (1) for Tevatron.}
\label{enogalhc(ii)}
\end{center}
\end{table*}

%*******************************************************

\section{Mixing in the Down Quark Sector: Phenomenology}

Let us now discuss how far the neutrino data can be useful in constraining
the RPV models in Scenario II with
mixing in the down-quark sector, $V=D_L=D_R$, $U_L=U_R={\bf 1}$. In this 
case the off-diagonal elements of the Yukawa coupling matrix $Y_d$ are
much larger than those in Scenario I (mixing restricted to the up quark 
sector only). As a result any input $\l'$-type couplings would lead to 
rather large values of $\kappa_i$ and $\l'_{i33}$ via RG evolution,  
in conflict with the
upper bounds in \cite{abada} unless the magnitudes of the input couplings
are too small to be of any phenomenological interest. This conclusion agrees
with that of \cite{allanach} although the latter was arrived at in a model 
with a single neutrino mass.

However, this argument hinges crucially on the RG evolution from the
GUT scale to the weak scale in the mSUGRA model. In practice,
the origin of SUSY breaking is not known. It is quite possible that SUSY
is broken at a much lower scale (as, {\em e.g.}, in the Gauge-mediated
SUSY breaking models).
For this type of models, not only the boundary conditions of the RG will
be different but as a whole the evolution should have a less important
role to play.

In view of the above uncertainties we take a more phenomenological point
of view and assume that only a few relatively large $\l'$ type couplings 
{\em in the weak basis}, not 
directly related to  $\nu$  physics, and bilinear parameters $\kappa_i$ are 
generated at the weak scale by some new physics at the high scale, the 
nature of which is unknown. We will see whether it is possible to
generate the suppressed $\l'_{i33}$ type couplings relevant for the 
$\nu$ sector via CKM rotation while the input  couplings at the weak 
scale can still be sufficiently large to trigger interesting non-neutrino
phenomenology.

In this case, the rotation to the mass eigenbasis is different 
from that in Scenario I. After rotation  eq.\ (\ref{superpot}) reads
\be
W_{L\!\!\!/} =
\tilde{\l}'_{imp}N_i D_mD^c_p + \bar{\l}'_{ijp}E_iU_jD^c_p,
\ee
where the fields are in the mass basis, and
\bea
\tilde{\l}'_{imp} &=& \l'_{ijk} V_{jm} V^\ast_{kp},\nonumber\\
\bar{\l}'_{ijp} &=& \l'_{ijk} V^\ast_{kp}.
    \label{ckm-opt1}
\eea
Thus in contrast to Scenario I new couplings involving both neutral and 
charged lepton superfields can be induced by the input couplings via CKM 
rotations. The next task is to find out the possible minimal sets of 
input couplings at the weak scale which can serve our purpose. 

Clearly, one can take $\l'_{i33}\not=0$ as the input,
each one with upper limits
of $1.5\times 10^{-4}$ approximately \cite{abada}. However, the induced 
couplings in this case will be so small that no interesting phenomenology
outside the neutrino sector apart from LSP decay \cite{barger}
and top squark decay \cite{shibu} is viable.  

Before we investigate other options, let us note that $\kkbar$ mixing data
predicts $|\l'_{i12}\l'_{i21}|< 10^{-9}$ \cite{gg-arc,jps-ak3}
(we drop the tilde from now, but these couplings are in the
mass basis). Thus, one
should not start with nonzero $\l'_{i11}$, $\l'_{i22}$, $\l'_{i12}$ or
$\l'_{i21}$ (in the weak basis) as inputs. From eq.\ (\ref{ckm-opt1}),
it is clear that one must start with such small values as to be devoid
of all phenomenological interests. The limiting value of all these
four sets is $1.4\times 10^{-4}$ at the weak scale \cite{abada}. 
This translates to
$|\l'_{i33}|<2.8\times 10^{-9}$, $2.5\times 10^{-7}$, and $2.6\times
10^{-8}$ (for both $\l'_{i12}$ and $\l'_{i21}$ sets) respectively. 
Considering the symmetric nature of the
magnitude of the CKM elements, one can start with either (i) $\l'_{i13(i31)}$  
or (ii) $\l'_{i23(i32)}$ to be nonzero. At first, the 
couplings are set to be nonzero at the weak scale with such a value as
to produce the limiting value of $1.5\times 10^{-4}$ for $\l'_{i33}$, and then
rescaled, if necessary, so as to conform with the $\Delta m_K$ constraint
(this is a similar iteration as was done in Section II). 

%-------------

Case (i): The maximum value of $\l'_{i23(i32)}$ to start with is $3.6
\times 10^{-3}$. After proper rotation, one obtains the couplings at the
mass basis and finds $|\l'_{i12}\l'_{i21}|<4.5\times 10^{-10}$, so this
choice is safe from the $\kkbar$ constraint. 

Case (ii): The neutrino data constrains $\l'_{i13(i31)}$ to be less than
$0.052$, but one needs a scaling down by a factor of $9.7$ to be in conformity
with the $\kkbar$ constraints.  For all these cases every other constraint
on product couplings is satisfied. 

However, with such small couplings, one does not hope to see an indirect
non-neutrino signal of RPV. 
We mention again that for indirect signals of
RPV, the individual couplings should be of the order of $10^{-2}$
and the product couplings of the order of $10^{-3}$-$10^{-4}$.
Such possibilities do not
exist for the option we have just discussed; but again, this is a
model-dependent statement and should not hold for other models of CKM 
mixing. 

Thus we conclude that irrespective of the details of the high scale
physics it is impossible to generate the $\l'_{i33}$ couplings of the 
benchmark scenario from relatively large, phenomenologically interesting 
input couplings at the weak scale and in the weak basis if 
(i) quark mixing is primarily restricted to the down quark
sector, and (ii) the Yukawa matrix is symmetric. 

Still, these couplings can induce
novel channels of decay for the lighter stop quark and the lightest
neutralino which are quite distinct for those in Scenario I.
Two possible phenomenological inputs couplings in the flavor basis at the weak
scale are  $\l'_{i31}$ or  $\l'_{i32}$. Since these couplings  themselves can
trigger top squark decays, the physical couplings in the mass basis does 
not
suffer any further suppression due to CKM rotation. The absence of bottom jets
in the final state distinguishes the top squark decay from that in Scenario I.
These signals have also been discussed in refs \cite{nkm,shibu}. 

\section{Conclusion}

In conclusion we reiterate that three relatively large input couplings of
$\l'_{ijk}$ type in the flavor basis at $M_G$ each bearing a different
lepton index can generate at the weak scale via RG evolution several
additional $\l'$-type couplings in the same basis (see Section III) with
naturally suppressed magnitudes. This occurs due to the flavor violation
inevitable in any model due to quark mixing. Also generated are three
lepton number violating bilinear terms $\kappa_i$. These bilinears
and three induced $\l'_{i33}$ couplings can explain the $\nu$   
oscillation data \cite{abada}. The upperbounds $\l'_{i33} < 1.5 \times
10^{-4}$ and $\kappa_i < 1.2\times 10^{-4}$ obtained from neutrino data can be
translated into upper bounds on the input couplings at $M_G$. In spite of
the severe constraints from the $\nu$ sector these input couplings are
allowed to be reasonably large.  When evolved to the weak scale and then
rotated to the mass basis the corresponding physical couplings can 
trigger several interesting 
signals, which are not expected in models where the couplings in
the $\nu$ sector are the only ones with appreciable magnitude.

In order to illustrate these ideas we study the RG equations of the RPV
mSUGRA model along with some models of the quark mixing matrix
\cite{agashe,allanach}. In models with quark mixing occuring in the up
sector (Scenario I)  we find that a minimal set of any
three couplings, chosen from the group $\l'_{i13}$ and $\l'_{i23}$, each
with a different lepton index can serve the purpose. For the input set
$\l'_{i13}$ (Case 1),  the oscillation data as well as other low
energy constraints \cite{rpv,gb}
lead to the upper bounds $\l'_{113}< 0.0064$, $\l'_{213}
< 0.019$, $\l'_{313} < 0.039$ at $M_G$ for $m_{\tilde b_R}=100$ GeV
(see Sections III and IV). Rare
$\tau$ decays with BRs close to the current experimental bounds are 
allowed in this case (see Table I). For the input set
$\l'_{i23}$ (Case 2), the bounds on the input couplings are more 
stringent, and hence no interesting rare weak process 
can be mediated by such small couplings. For the mixed choice 
$\l'_{123}$, $\l'_{213}$ and $\l'_{313}$, the rare decay $K\to\pi\nu\bar{\nu}$
with BR close to the current experimental upper limit  are consistent with
the bounds on the input couplings.    

LSP decays with characteristic BRs occur for all choices of GUT scale
inputs. For Case 1 LSP decays into a charged lepton accompanied by a
$b$-jet and a light quark jet is the distinctive signal (Table II) . In
Case 2 two heavy flavor jets ($c$ and $b$) accompany the
charged lepton (Table III). The relative abundance of the final states 
with different leptons can indicate the relative strengths of the inputs 
at $M_G$.

The lighter top squark decays in RPV channels can naturally compete 
with its RPC decays in models of $\nu$ mass provided this sparticle 
happens to be the NLSP \cite{nkm,shibu}. In Case 1, due to the relatively
large upper bounds on the input RPV couplings, RPV can overwhelm RPC 
decays in a large region of the parameter space. However, if the input 
couplings (in particular the one involving $\tau$)  are somewhat 
smaller than the upper bounds, competition between RPV and RPC modes may 
show up (Table VII).  Competition among the  RPV and RPC decays are 
more probable in Case 2 (Table VIII).

In Case 1 the upper bounds on the RPV couplings are consistent with 
direct RPV decays of other squarks and leptons with sizable BRs
(Tables IV-VI).
The above decays can lead to several low background signals both at the 
Tevatron and at the LHC. At Tevatron gaugino pair production followed by 
the LSP decay can lead to interesting signals (Tables XV and XVI).    
 
Top squark pair production at Tevatron followed by direct RPV decays of 
both the sparticles into an OSD pair of same flavor and jets 
have been discussed in \cite{nkm,shibu}. In this paper we consider
the signals with isolated LSD pairs and isolated dileptons of different 
flavors (Tables IX and X corresponding to Case 1 and Case 2, respectively). 
Both are expected to be low background signals and our numerical estimates 
indicate large number of events. Obviously a huge signal at LHC is anticipated.

Gaugino pair production followed by LSP decays can provide the signals 
similar to the ones discussed in the last paragraph. Even at the Tevatron
a healthy size of the signal is expected in both Case 1 and 2 (Tables
XIII and XIV). The prospect of observing these signals at the LHC is even 
better (Tables XV and XVI).

Interesting signals arising from sfermion-antisfermion  pair production
followed by the direct RPV decay of one and indirect RPV decay of the other
can lead to signals with decent magnitudes at LHC (Tables XI and XII).     

If, on the other hand, a model of the CKM matrix with mixing in the down
quark sector (Scenario II) is considered, an mSUGRA type model with 
phennomenologically interesting, large input RPV couplings 
at $M_G$ not directly related to the $\nu$ sector is not viable.
If attention is paid to the $\nu$ constraint the input couplings at 
$M_G$ are destined to be highly suppressed (Section VI)

We next consider a phenomenological model with relatively large 
weak scale RPV parameters in the flavor basis induced by some high scale 
physics, the nature of which is presently unknown. Even if the induced 
trilinear couplings 
are not directly related to the neutrino sector, the $\l'_{i33}$
couplings can be generated in the mass basis via CKM rotations. 
One can start with either (i) $\l'_{i13(i31)}$  
or (ii) $\l'_{i23(i32)}$ to be nonzero in the flavor basis at the weak 
scale. Unfortunately CKM rotations of these inputs not only generate
the couplings required by the $\nu$ sector consistent with their 
upper limits, but also generate other
dangerously large couplings leading to unaccetably large $\kkbar$
mixing amplitude. Consequently such inputs must have small magnitudes 
not interesting for rare weak decays or direct RPV decays of squarks or 
sleptons. However top squark and LSP decays occur with characteristics 
quite different from the ones discussed for Scenario I (Section VI).          

\section{Acknowledgements}

AK was supported by the project no.\ SR/S2/HEP-15/2003 of DST, Govt.\
of India.

%*****************************************************************

\end{document}